\documentclass[conference]{IEEEtran}
\IEEEoverridecommandlockouts

\usepackage[
 colorlinks=true,
 allcolors=black,
 bookmarks,
 unicode
]{hyperref}

\usepackage{xspace}
\usepackage{pgfplotstable}
    \pgfplotsset{
        compat=1.7,
    }
\usepackage{adjustbox}
\usepackage{booktabs} 
\usepackage{pifont}
\usetikzlibrary{shapes,snakes,arrows,automata}
\usetikzlibrary{positioning}
\usepackage{savesym}
\usepackage{fourier}
\usepackage[clock]{ifsym}
\usepackage{listings}
\usepackage{fontawesome}
\usepackage{inconsolata}
\usepackage{tcolorbox}
\usepackage{amsmath}
\usepackage{csquotes} 
\usepackage{enumitem}

\usepackage{microtype} 

\definecolor{mygreen}{rgb}{0,0.5,0}
\definecolor{myred}{HTML}{CC2222}
\definecolor{cobalt}{rgb}{0.0, 0.28, 0.67}
\DeclareMathSymbol{\qm}{\mathalpha}{operators}{"3F}
\newtcolorbox{mybox}[2][]{title=#2,#1,colframe  = black!100,}

\newcounter{bulletscount}

\def\Snospace~{\S{}}

\lstdefinestyle{CStyle}{
    frame=none,
    keywordstyle=\color{cobalt},
    language=C,
    classoffset=1, 
    otherkeywords={uint8_t,,uint32_t},
    morekeywords={uint8_t,uint32_t},
    keywordstyle=\color{cobalt},
    classoffset=2, 
    otherkeywords={ridx,x,y,temp,k, idx, case_13, load_value},
    morekeywords={ridx,x,y,temp,k, idx, case_13, load_value},
    keywordstyle=\color{violet},
    classoffset=3, 
    otherkeywords={A,B,C},
    morekeywords={A,B,C},
    keywordstyle=\color{mygreen},
}
\lstdefinestyle{x86Style}{
    frame=none,
    keywordstyle=\ttfamily\bf,
    language=C,
    morekeywords={mov, sub, and, movzx, ret, push, call, add},
    classoffset=1, 
    otherkeywords={temp, case_13, load_value},
    morekeywords={temp, case_13, load_value},
    keywordstyle=\color{violet},
    classoffset=3, 
    otherkeywords={A,B,C},
    morekeywords={A,B,C},
    keywordstyle=\color{mygreen},
}
\def\fullbox[#1,#2,#3,#4,#5]#6{
	\draw node[draw,color=gray!50,minimum
	height=#1,minimum width=#2] (#4) at #5 {}; 
	\node[anchor=#3,inner sep=2pt] at (#4.#3)  {#6};
}
\tikzstyle{fancytitle} =[draw,fill=white, text=black, rounded corners=2, inner sep=2pt]
\newcolumntype{Y}{>{\centering\arraybackslash}X}
\newcolumntype{Z}{>{\raggedright\arraybackslash}X}
\newcolumntype{C}[1]{>{\raggedright\arraybackslash}p{#1}}
\tcbset{tab2/.style={enhanced,fonttitle=\bfseries,fontupper=\normalsize\sffamily,
colback=yellow!10!white,colframe=red!50!black,colbacktitle=red!40!white,
coltitle=black,center title}}
\newlength\ideawidth

\hypersetup{
  pdftitle={Cats vs. Spectre: An Axiomatic Approach to Modeling Speculative Execution Attacks},
  pdfauthor={Hernán Ponce-de-León, Johannes Kinder},
  pdfkeywords={security, speculative execution, spectre, axiomatic semantics},
}

\newcommand\regs{\it{Regs}}

\newcommand\nat{\mathbb{N}}

\newcommand\commited{\mathbb{C}}
\newcommand\executed{\mathbb{X}}

\newcommand\transient{\mathbb{T}}

\newcommand\stores{\mathbb{W}}
\newcommand\loads{\mathbb{R}}
\newcommand\memory{\mathbb{M}}

\newcommand\iskip{\textbf{skip}\xspace}
\newcommand\load{\textbf{load}\ }

\newcommand\store{\textbf{store}\ }
\newcommand\jmp{\textbf{jmp}\ }
\newcommand\lb{\ell}
\newcommand\lbs{\nat}
\newcommand\beqz{\textbf{beqz}\ }
\newcommand\fence{\textbf{fence}\xspace}
\newcommand\secret{\textbf{sec}}
\newcommand\red{\color{myred}}
\newcommand\mygreen{\color{mygreen}}

\newcommand\pred[1]{\textsc{pred}(#1)}

\newcommand\cfd[2]{\textsc{cfd}(#1,#2)}
\newcommand\scfd[2]{\textsc{scfd}(#1,#2)}

\newcommand\com[1]{e_{#1} \in \commited}

\newcommand\cf[1]{e_{#1} \in \executed}

\newcommand\scf[1]{e_{#1} \in \transient}

\newcommand\cp[1]{\texttt{bpc}_{e_{#1}}}

\newcommand\window{\texttt{sw}\xspace}

\newcommand\rval[1]{\ensuremath{\mathit{val}({#1})}}
\newcommand\reg[1]{\ensuremath{\mathit{reg}({#1})}}
\newcommand\expr[1]{\ensuremath{\mathit{exp}({#1})}}

\newcommand\define{\mathrel{:=}}

\newcommand\varA{{\color{mygreen} \texttt{A}}\xspace}
\newcommand\varB{{\color{mygreen} \texttt{B}}\xspace}
\newcommand\varC{{\color{mygreen} \texttt{C}}\xspace}
\newcommand\varx{{\color{violet} \texttt{x}}\xspace}
\newcommand\vary{{\color{violet} \texttt{y}}\xspace}
\newcommand\varidx{{\color{violet} \texttt{idx}}\xspace}

\newcommand\vartemp{{\color{violet} \texttt{temp}}\xspace}

\newcommand\masm{$\mu$\textsc{Asm}\xspace}

\newcommand\gtick{{\color{mygreen}{\ding{52}}}}

\newcommand\unknown{\textsc{Unknown}\@\xspace}
\newcommand\safe{\textsc{Safe}\@\xspace}
\newcommand\unsafe{\textsc{Unsafe}\@\xspace}
\newcommand\sat{\textsc{Sat}\@\xspace}
\newcommand\unsat{\textsc{Unsat}\@\xspace}

\newcommand\tso{\textsc{TSO}\@\xspace}
\newcommand\arm{\textsc{ARM}\@\xspace}
\newcommand\armv{\textsc{ARM}v8\@\xspace}
\newcommand\power{\textsc{Power}\@\xspace}
\newcommand\celeven{C11\@\xspace}

\newcommand\dartagnan{\textsc{Dartagnan}\@\xspace}
\newcommand\zombmc{\textsc{Kaibyo}\@\xspace}

\newcommand\pitchforktool{\textsc{Pitchfork}\@\xspace}
\newcommand\impro{\texttt{IMP-ro}\@\xspace}

\newcommand\spectector{\textsc{Spectector}\@\xspace}
\newcommand\sklee{\textsc{KLEESpectre}\@\xspace}
\newcommand\binsec{\textsc{Binsec}\@\xspace}
\newcommand\checkmate{\textsc{CheckMate}\@\xspace}

\newcommand\zthree{\textsc{Z3}\@\xspace}
\newcommand\yices{\textsc{Yices2}\@\xspace}
\newcommand\mathsat{\textsc{MathSAT5}\@\xspace}
\newcommand\cvcfour{\textsc{CVC4}\@\xspace}

\newcommand\xes{\textsc{x86}\@\xspace}
\newcommand\spectre {\textsc{Spectre}\@\xspace}

\newcommand\fen {\textsc{Fence}\@\xspace}
\newcommand\none {\textsc{None}\@\xspace}

\newcommand\flushreload {\textsc{Flush+Reload}\@\xspace}
\newcommand\primeprove {\textsc{Prime+Probe}\@\xspace}

\newcommand{\relname}[1]{\textsf{\small#1}}

\newcommand\bnrel{\langle \mathit{rel} \rangle}
\newcommand\bnass{\langle \mathit{assert} \rangle}
\newcommand\bnev{\langle \mathit{set} \rangle}
\newcommand\bntermbase{\langle \mathit{b} \rangle}
\newcommand\bnterm{\langle \mathit{r} \rangle}
\newcommand\bnmcm{\langle \mathit{MCM} \rangle}
\newcommand{\name}{\langle\mathit{name}\rangle}

\newcommand\bnexp{\langle \mathit{exp} \rangle}
\newcommand\bnstm{\langle \mathit{stm} \rangle}
\newcommand\bnprogram{\langle \mathit{p} \rangle}

\newcommand\tot[2]{\mathit{total}(#1,#2)}

\newcommand\acy[1]{\mathit{\color{cobalt}\relname{acyclic}} #1}

\newcommand\irref[1]{\mathit{\color{cobalt}\relname{irreflexive}} #1}
\newcommand\empt[1]{\mathit{\color{cobalt}\relname{empty}} #1}

\newcommand{\johannes}[1]{\textbf{\color{red}JK: #1}}

\begin{document}

\title{Cats vs. Spectre: An Axiomatic Approach to Modeling Speculative Execution Attacks}

\author{\IEEEauthorblockN{Hern\'an Ponce-de-Le\'on}
\IEEEauthorblockA{Research Institute CODE\\Bundeswehr University Munich\\hernan.ponce@unibw.de}
\and
\IEEEauthorblockN{Johannes Kinder}
\IEEEauthorblockA{Research Institute CODE\\Bundeswehr University Munich\\johannes.kinder@unibw.de}
}

\maketitle
\thispagestyle{plain}
\pagestyle{plain}

\begin{abstract}
The \spectre{} family of speculative execution attacks has required a rethinking of formal methods for security. Approaches based on operational speculative semantics have made initial inroads towards finding vulnerable code and validating defenses. 
However, with each new attack grows the amount of microarchitectural detail that has to be integrated into the underlying semantics. 
We propose an alternative, light-weight and axiomatic approach to specifying speculative semantics that relies on insights from memory models for concurrency. We use the CAT modeling language for memory consistency to specify execution models that capture speculative control flow, store-to-load forwarding, predictive store forwarding, and memory ordering machine clears.
We present a bounded model checking framework parameterized by our speculative CAT models and evaluate its implementation against the state of the art. Due to the axiomatic approach, our models can be rapidly extended to allow our framework to detect new types of attacks and validate defenses against them.
\end{abstract}

\section{Introduction}
\label{sec:introduction}

The promise of formal methods is to provide guarantees of correctness and security in exchange for rigorous specifications and sound layering of abstractions~\cite{ChongGDMPSSZ16}. 
Unfortunately, the discovery of speculation-based vulnerabilities such as \spectre has shown that some of the most basic abstractions are broken at the microarchitectural level in today's mainstream computing architectures~\cite{KocherHFGGHHLM019}. 
Modern processors execute code speculatively and may have to roll back state if a prediction turns out to be wrong; despite being rolled back, the transient execution can leave traces in the microarchitectural state that an attacker can abuse. By causing the processor to mispredict certain conditions, such as whether a branch is taken or not, an attacker can exfiltrate data despite system-level protection. 

There is ongoing work to incorporate microarchitectural effects into formal methods, in order to resolve this impasse~\cite{CauligiDGTSRB20,DanielBR20,GuancialeBD20,GuarnieriKMRS20,spectaint,WangCBMR20,oo7,Wu019}. The proposal is appealing: if the semantics used for reasoning about code takes speculation into account, we can reliably identify code patterns vulnerable to transient execution attacks, and we can prove the effectiveness of proposed countermeasures. 
A key challenge here is the variety of attacks: while commonly referred to under the umbrella term of \emph{speculative execution attacks}, the underlying mechanisms can be very different, and require different semantics.

The majority of approaches rely on defining speculative \emph{operational} semantics.
Operational semantics describe \emph{how} a valid program is interpreted, as sequences of computational steps.
This requires descriptions of microarchitectural implementation details such as buffers or recovery mechanisms for wrong predictions. 
For example, operational semantics for speculative execution record snapshots of the state before every prediction to enable subsequent rollbacks.
This results in complex models and complications with nested speculations, which have been listed as one of the main challenges of modeling speculative execution~\cite{GuarnieriKMRS20,OleksenkoTSF20}.
Porting an operational speculative semantics to incorporate a different class of attack is no easy task, and no such approach covers all known attacks. 

\emph{Axiomatic} semantics, as an alternative to the operational approach, define \emph{which} executions are valid. The axiomatic approach has been successful in reasoning about concurrency and weak memory models. Its elegance lies in being able to succinctly express which dependencies among reads and writes are enforced. 
This enables a modular style of reasoning.
To describe different memory models, Alglave et al.~introduced the relational language \emph{CAT}, which is expressive enough to axiomatize the concurrency semantics of processors like \xes, \power and \arm~\cite{cat,AlglaveDGHM21,AlglaveMT14}.

We argue that the axiomatic approach and CAT in particular lend themselves equally well to modeling speculative execution semantics. To this end, we develop a set of models describing speculative behaviors and their effects. 
We find CAT to be ideally suited to capture a variety of attacks in a simple, concise and unified manner. 
While the similarity between weak memory models and speculation has been noted before~\cite{ColvinW19,DisselkoenJJR19,checkmate}, we are the first to prove their flexibility by providing axiomatic weak memory-style models for several variants of speculative execution attacks and a concrete tool to detect vulnerable code and validate defenses. In a recent survey, Cauligi et al.~\cite{sok:spectre} stated that \enquote{\emph{they are only suited for analyzing particular Spectre variants [$\ldots$] and are difficult to adapt to other attacks}} and that \enquote{\emph{it remains an open problem to translate a semantics of this style into a concrete analysis tool}}.

The key advantages of defining axiomatic semantics with CAT are \emph{simplicity} and \emph{modularity}, which lead to reliable and rapid development of verification tools.
Because the models are simple and concise, the resulting analysis tools are less involved. 
Because the speculative semantics defined as CAT models are modular, we can quickly adapt to new types of attacks. A case in point is that in the final phase of preparing this paper, Ragab et al.~\cite{ragab_rage_2021} presented machine clears as a new source for transient execution. Following the paper's description, we were able to define a CAT model to handle the memory ordering machine clear in less than two hours (see~\autoref{sec:concurrency}).

Overall, this paper makes the following contributions:
\begin{itemize}
  \item We show how to use the CAT language to axiomatically describe the effects of microarchitectures and provide concrete models that capture the behaviors underlying known \spectre attacks: speculative control flow (\autoref{sec:semantics}), store-to-load forwarding, predictive store forwarding, and memory ordering machine clear (\autoref{sec:cat-models}).
  \item We propose an analysis framework that is parametric in its microarchitectural model (defined via CAT). The analysis is an instance of a Bounded Model Checking problem, implemented in the tool \zombmc, which we use to verify software isolation (\autoref{sec:framework}).
  \item We evaluate our framework and compare its precision, flexibility, and performance against state-of-the-art tools to detect \spectre vulnerabilities (\autoref{sec:evaluation}).
\end{itemize}

\section{Preliminaries}
\label{sec:preliminaries}

We begin by introducing the target assembly language (\autoref{sec:language}) and its semantics (\autoref{sec:exec}). 
We then present the CAT language (\autoref{sec:cat}), followed by a brief description of speculative execution attacks (\autoref{sec:se}), the threat model we consider (\autoref{sec:threatmodel}), and possible extensions (\autoref{sec:beyond-isolation}).

\subsection{Input Language}
\label{sec:language}

\begin{figure}[t]
\setlength{\tabcolsep}{3pt}
\scalebox{.9}{
$$\begin{tabular}{lccl}
\textbf{Basic Types} \\
(Registers) & $r$ & $\in$ & \regs \\
(Values)    & $n$ & $\in$ & $\nat$ \\ 
\\
\textbf{Syntax} \\
(Expressions) & $\bnexp$ & $\define$ & $r \mid n \mid \ominus \bnexp \mid \bnexp \otimes \bnexp \mid \secret$ \\
(Statements)  & $\bnstm$   & $\define$ & $r \leftarrow \bnexp \: \mid \: r \stackrel{\bnexp ?}{\longleftarrow} \bnexp$ \\
      &   & $\mid$  & $\load r,\bnexp \: \mid \: \store r,\bnexp$ \\
      &   & $\mid$  & $\jmp \lb \: \mid \: \beqz r,\lb$ \\
      &   & $\mid$  & $\iskip \: \mid \: \fence$ \\
\\
(Labels)    & $\lb$ & $\in$ & $\lbs$ \\
(Programs)  & $\bnprogram$ & $\define$ & $\lb : \bnstm \mid \bnprogram ; \bnprogram$ \\
\end{tabular}$$}
\caption{\masm syntax.}
\label{fig:masm}
\end{figure}

We target \masm, a core assembly language defined by Guarnieri et al.~\cite{GuarnieriKMRS20}. The syntax is given in \autoref{fig:masm}.
Registers and constants over $\nat$ form the base expressions;
more complex expressions are built using the usual unary and binary operators.
Besides computing values, expressions are used as memory addresses.
We introduce the special expression $\secret$ to represent the address of a secret (more on this in \autoref{sec:threatmodel}).

The statements of the \masm language are (conditional) local assignment, memory \textbf{load} and \textbf{store}, direct (\textbf{jmp}) and conditional (\textbf{beqz}) jump, \textbf{fence} and \textbf{skip}.
A statement is paired with a unique label $\lb \in \lbs$.
The resulting pair $\lb : s$ is called an instruction.
We refer to instructions by their labels.
We use $\reg{\lb}$ and $\expr{\lb}$ to refer to the target register and the memory expression of a given instruction, respectively.
Finally, programs are sequences of instructions.

Statement $r \leftarrow \mathit{e}$ computes the value of expression $\mathit{e}$ and locally assigns it to register~$r$.
The conditional assignment \smash{$r \stackrel{\mathit{e}' ?}{\longleftarrow} \mathit{e}$} takes effect only if $\mathit{e}'$ does not evaluate to 0.
Statement $\load r, \mathit{e}$ assigns the value at the address computed from $\mathit{e}$ to register $r$.
Conversely, $\store r, \mathit{e}$ assigns the value of $r$ to the address computed from $\mathit{e}$.
Statement $\jmp \lb$ directly redirects the control flow to $\lb$.
The conditional jump $\beqz r,\lb$ redirects the control flow to $\lb$ only if the value of register $r$ evaluates to 0.
A \fence enforces an ordering constraint on memory operations.
A \iskip statement has no computational effect.
We define the set of static predecessors of label $\lb$ as
$$\begin{tabular}{lcl}
${\pred \lb}$ & $\define$ & $\big\{ \lb' \in \lbs \mid {({\lb'+1 = \lb} \land {\lb' \neq \jmp \lb''})} \lor {}$ \\
& & $\hspace{22mm} {\lb' = \jmp \lb} \lor {\lb' = \beqz r,\lb} \big\}.$
\end{tabular}$$
In slight abuse of notation, we write $\lb = \jmp \lb'$ here to mean that $\lb : \jmp \lb'$ is an instruction contained in the program. As a result of the definition, ${\pred \lb}$ contains the instruction immediately preceding $\lb$ in the program order, if that is not a \jmp instruction, and any jump (direct or conditional) in the program targeting $\lb$.

\begin{figure}[t]
\centering
\scalebox{.8}{

\begin{tikzpicture}

  \node(exec) at (0,0) {\color{black!60}\textit{=== Candidate Execution ===}};  
  \node (n0) [below=1mm of exec] {$e_1: \load r_1, \varidx$};
  \node (n1) [below of=n0] {$e_2: r_2 \leftarrow (r_1 < \varA.\texttt{size})$};
  \node (n2) [below of=n1] {$e_3: \beqz r_2, e_7$};
  \node (n3) [below of=n2] {$e_4: \load r_3, \varA + r_1$};
  \node (n4) [below of=n3] {$e_5: \load r_4, \varB + r_3$};
  \node (n5) [below of=n4] {$e_6: \vartemp \leftarrow \vartemp\ \&\ r_4$};
  \node (n6) [below of=n5] {$e_7: \bf skip$};
  
  \node (secret)[left=1cm of n3] {$e_s: \secret$};

  \node(C0) at (-7,0) {\color{black!60}\textit{=== C code ===}};  
  \node(C) [below=.1mm of C0.south west,anchor=north west] {%
  \begin{lstlisting}[style=CStyle]
if (idx < A.size) {
  temp &= B[A[idx]];
}
  \end{lstlisting}%
  };

  \node(i0) [below=13.4mm of C.west,anchor=west]{\color{black!60}\textit{=== \masm code ===}};
  \node(i1) [below=.6cm of i0.west,anchor=west]{1: $\load r_1, \varidx$};
  \node(i2) [below=.6cm of i1.west,anchor=west]{2: $r_2 \leftarrow (r_1 < \varA.\texttt{size})$};
  \node(i3) [below=.6cm of i2.west,anchor=west]{3: $\beqz r_2, 7$};
  \node(i4) [below=.6cm of i3.west,anchor=west]{4: $\load r_3, \varA + r_1$};
  \node(i5) [below=.6cm of i4.west,anchor=west]{5: $\load r_4, \varB + r_3$};
  \node(i6) [below=.6cm of i5.west,anchor=west]{6: $\vartemp \leftarrow \vartemp\ \&\ r_4$};
  \node(i7) [below=.6cm of i6.west,anchor=west]{7: $\bf skip$};
    
  \path
	(n0) edge[->]			node[left] {\relname{po}} 		(n1)
	(n1) edge[->]			node[left] {\relname{po}} 		(n2)
	(n2) edge[->]			node[left] {\relname{po}} 		(n3)
	(n3) edge[->]			node[left] {\relname{po}} 		(n4)
	(n4) edge[->]			node[left] {\relname{po}} 		(n5)
	(n5) edge[->]			node[left] {\relname{po}} 		(n6);
  
  \path
	(secret) edge[->, dashed]			node[above] {\relname{rf}}		(n3);  
  \path
	(n2) edge[->, bend right=-40, dashed]			node[right] {\relname{fence}}		(n3);  

\end{tikzpicture}}
\caption{\spectre-v1 -- impossible under traditional semantics, but possible under control flow speculation.}
\label{fig:pht}
\end{figure}

Consider the left of~\autoref{fig:pht} which shows a code snippet written in C (top) and \masm (bottom).
Variable \varidx is an input and \varA.\texttt{size} is the length of the array \varA.
If $r_2 \not = 0$ (meaning that \varidx is in bounds for \varA), instruction 4 loads the value of \texttt{{\varA}[{\color{violet}idx}]} (here represented by accessing address $\varA + r_1$) into register $r_3$; if not, the program terminates by jumping to 7.
Since the input is compared with the size of the array, instruction 4 cannot read from arbitrary memory. In particular, it cannot access the secret at address \secret.
The loaded value in $r_3$ is used by instruction 5 for accessing a second array \varB.
For simplicity, we assume the values of \varA are smaller than the size of \varB, so that there is no need for a second bounds check. We then can conclude that the program cannot access memory out of bounds.

\subsection{CAT Semantics of \masm}
\label{sec:exec}

We define the semantics of a program axiomatically in terms of its \emph{consistent executions}, following the CAT approach introduced by Alglave et al.~\cite{cat,AlglaveMT14} for formalizing weak memory models. 
Behaviors of a program are represented by graphs where nodes (called events) model occurrences of instructions and edges model relations or dependencies.
Given a program, we proceed in two steps.
First, we define all possible behaviors or \emph{candidate executions}, which satisfy basic properties about control and data flow.
Second, we filter out behaviors that are invalid with respect to the target semantics, using a set of assertions given as a CAT model. 
The remaining behaviors form the set of consistent executions and define the semantics of the program.
Different assertions yield different CAT models, each of them describing one concrete semantics. Note that for a given program, different assertions might result in the same set of consistent executions. When this is true for every possible program, the CAT models are equivalent. 
Our CAT model for in-order semantics is given in~\autoref{fig:stl-safe}.

An \emph{event} is the representation of an occurrence or instance of an executed instruction.
Let $\executed$ be a set of events representing a behavior, i.e., the nodes of the corresponding graph.
There are certain properties $\executed$ must fulfill to guarantee that the behavior represents a possible control flow of the program.
Since, on top of traditional control flow, we model speculative execution, we leave such properties underspecified here; they are formalized in~\autoref{sec:semantics}.
For loop-free programs there is a one to one correspondence between events and executed instructions: there is a unique instance of $\lb$ represented by $\cf{\lb}$.
In the presence of loops, if $\lb$ is in a loop that is executed $n$ times, then there are $n$ such instances and $\{ e^1_\lb, \dots, e^n_\lb \} \subseteq \executed$.
Events coming from memory instructions form the set $\memory \subseteq \executed$, which is further split into $\loads$ and $\stores$ depending on whether instructions come from \load or \store statements, i.e., $\memory = \loads \uplus \stores$.
By $\memory_a$ we refer to memory events that access an address $a \in \nat$.
Finally, if $\lb$ is an instruction using a register and $e_\lb$ is an event representing an instance of $\lb$, we use $\rval{e_\lb}$ to represent the value of the register for that given instance.

Relations form the edges of execution graphs.
The \emph{location relation} \relname{loc} forms equivalence classes between memory events accessing the same address, i.e., $\relname{loc} \define \left\{ (e_\lb,e_{\lb'}) \mid \exists a \in \nat : e_\lb, e_{\lb'} \in \memory_a \right\}$.
The \emph{reads-from relation} $\relname{rf}$ gives for each read a unique write to the same address from which the read obtains its value:
\begin{gather*}
\relname{rf} \subseteq (\stores \times \loads) \cap \relname{loc}\\
\forall r \in \loads : {\exists! w \in \stores : {\relname{rf}(w,r)}}\\
\relname{rf}(w,r) \Rightarrow \rval{w} = \rval{r}
\end{gather*}
The last constraint above defines how the data flows between different instructions.
The candidate execution of~\autoref{fig:pht} represents a behavior where the initial value of the secret (here represented by $e_s$) flows to the access to \varA at index $r_1$; this is represented by the edge $\relname{rf}(e_s,e_4)$.

The \emph{coherence order} $\relname{co}$ relates writes to the same address and forms a total order for each address:
\begin{gather*}
\relname{co} \subseteq (\stores \times \stores) \cap \relname{loc}\\
\forall a \in \nat : \tot{\relname{co}}{\stores_{a}}
\end{gather*}
Coherence models the order in which \store instructions hit the main memory.
For each address, we assume the existence of a write event assigning its initial value. 
Those events come first in the coherence order.  
In figures we represent initial writes to all addresses by a unique event $e_0: \mathit{init}$.
The only exception is address \secret\ for which we use $e_s:\secret$; this event models the initial value of the secret.

The \emph{program order} $\relname{po} \subseteq (\executed \times \executed)$ represents the order in which instructions are written.
For instructions not being part of the same loop we have 
$(e_\lb, e_{\lb'}) \in \relname{po} \Rightarrow \lb < \lb'.$
For instructions belonging to the same loop, we have that 
$$(e^i_\lb, e^j_{\lb'}) \in \relname{po} \Rightarrow {({\lb < \lb'} \land {i \leq j})} \lor {({\lb \geq \lb'} \land i < j)}.$$
In~\autoref{fig:pht} we have \relname{po} edges between all events $e_1$-$e_7$; they represent the order of the corresponding instructions  in the \masm code.
Relation $\relname{fence}$ contains every pair of events for which there is a \fence instruction in between, i.e.,
$\relname{fence} \define \left\{(e_\lb,e_{\lb'}) \subseteq \relname{po} \mid \exists \lb'':\fence : \lb < \lb'' < \lb' \right\}$.
Note that relations \relname{co}, \relname{po} and \relname{fence} are transitive, although we represent them in diagrams only by their direct edges.
The \emph{address dependency} \relname{addr} relates reads with memory events using the loaded value for computing their addresses:
\addtolength{\tabcolsep}{-4pt}
$$\begin{tabular}{lcl}
\relname{addr} & $\define$ & $\big\{(e_\lb,e_{\lb'}) \subseteq (\loads \times \memory) \cap \relname{po} \mid \reg{\lb} \in \expr{\lb'}$ \\
& & $\hspace{18mm} {} \land \not \exists \lb'': \lb < \lb'' < \lb' \land \reg{\lb''} = \reg{\lb} \big\} $ \\
\end{tabular}$$

A candidate execution is a triple $(\executed, \relname{rf}, \relname{co})$.
Each different combination of these three yields a possible behavior.
Once $\executed$ is fixed, relations \relname{po}, \relname{fence} and \relname{addr} can be statically computed from the program.
The right part of~\autoref{fig:pht} shows one candidate execution for the program on the left (note that certain \relname{rf} edges are omitted for clarity).

\subsection{Specifying Memory Models with CAT}
\label{sec:cat}

Each candidate execution as defined in~\autoref{sec:exec} yields a possible behavior.
On a given processor, however, only some of those behaviors can occur in practice.
A memory model defines which behaviors are allowed, by specifying which values a \load instruction can read.
We use CAT, the core of which is given in~\autoref{fig:model}, to formalize this.
CAT is concise but expressive enough to specify a wide range of memory models. 
It has been used to clarify and formalize the concurrency semantics not only of processors like \xes, \power and \arm, but also the memory model of \celeven and the Linux kernel~\cite{AlglaveMMPS18, AlglaveMT14, BattyDW16, PulteFDFSS18, SarkarSNORBMA09}.
Moreover, the CAT model of \armv is part of its official documentation~\cite{AlglaveDGHM21}.
In this paper we show that CAT can also be used to model the effects of microarchitectural optimizations such as branch and alias predictors.

The role of the memory model is to filter out the candidate executions that are not consistent according to the intended semantics.
In CAT, a memory model is a constraint system over so-called \emph{derived relations}. 
Derived relations are built from the base relations described in~\autoref{sec:exec}, hand-defined relations that refer to the different sets of events, and named relations that we will explain in a moment.
CAT supports operators like union, intersection, difference, inverse, transitive (and reflexive) closure and composition.
The assertions that filter candidate executions are acyclicity, irreflexivity and emptiness constraints over derived relations.
As an example, our CAT model for in-order semantics in~\autoref{fig:stl-safe} defines the derived relation \relname{com} as the union of \relname{co}, \relname{rf} and the composition of \relname{rf}$^{-1}$ with \relname{co}.
The model states that the union of \relname{com} and \relname{po} must be acyclic.
CAT also supports recursive definitions of relations.
We assume a set $\name$ of relation names (different from the predefined relations) and require each name used in the memory model to have a defining equation $\name=\bnterm$. 
Notably, $\bnterm$ may again contain named relations, making the system of defining equations recursive.
The relations then are defined to be the least solution to this system of equations.

\begin{figure}[t]\small
		\begin{align*}
		\bnmcm  \define &\  \bnass\mid \bnrel \mid \bnmcm \wedge \bnmcm\\
		\bnass  \define &\ \acy \bnterm  \mid \irref \bnterm \mid \empt \bnterm  \\
		\bnterm \define&\  \bntermbase \mid \bnterm \cup \bnterm \mid \bnterm \cap \bnterm \mid \bnterm \setminus \bnterm \\
		& \mid \bnterm^{-1}  \mid \bnterm^+ \mid \bnterm^* \mid \bnterm;\bnterm  \\
		\bntermbase \define &\ \relname{po}  \mid \relname{fence} \mid \relname{rf} \mid \relname{co} \mid \relname{loc} \mid \relname{addr} \\
		& \mid [\bnev] \mid \bnev \times \bnev\mid \name  \\
		\bnev  \define &\ \executed \mid \memory \mid \stores \mid \loads\\
		\bnrel  \define &\ \name = \bnterm
		\end{align*}
		\caption{Core of the CAT language~\cite{cat}.}
		\label{fig:model}
\end{figure}

\subsection{Speculative Execution Attacks}
\label{sec:se}

Speculative execution uses different predictors to guess, e.g., the outcome of branching instructions or aliasing of addresses.
A prediction opens a \emph{speculation window} during which instructions are executed without knowing whether the prediction was correct. 
Once the window closes, the effects of the speculatively executed instructions are committed (if the prediction was correct) or rolled back (if the prediction was wrong).

Speculatively executed instructions under a misprediction are called \emph{transient}.
When speculation is over, all directly visible effects of transient instructions are discarded.
However, this is only the case for the architectural state.
If the cache or other microarchitectural components have been modified by these instructions, those effects are not (or only partially) discarded.
This can allow sensitive data to be leaked, e.g., through cache timing~\cite{KocherHFGGHHLM019,LiuYGHL15}.
In these attacks, the size of the speculation window determines the number of operations an attacker can issue on a transient path before the results are squashed.

The candidate execution in~\autoref{fig:pht} represents a \spectre-v1 gadget~\cite{KocherHFGGHHLM019}.
An attacker can bypass the bounds check in the following way~\cite{CanellaB0LBOPEG19}:
first, during the \emph{setup phase}, the attacker invokes the code with valid values of \varidx, thereby training the branch predictor to expect the jump not to be taken.
Second, during the \emph{transient execution phase}, the attacker invokes the program with a value of \varidx outside the bounds of \varA.
Rather than waiting for the result of the comparison (e.g., if the value of \varA.\texttt{size} is not cached), the processor guesses that the bounds check will be true and transiently executes instructions 4 and 5 using the malicious \varidx.
Note that instruction 5 loads data into the cache in an address that is dependent on \texttt{{\varA}[{\color{violet}idx}]}.
When the result of the bounds check is eventually determined, the processor rolls back the effect of 4 and 5. 
However, changes made to the cache state are not reverted.
Finally, in the \emph{decoding phase}, the attacker can use a side channel to analyze the cache contents and retrieve the value of the secret~\cite{LiuYGHL15}.

\subsection{Threat Model}
\label{sec:threatmodel}

An attacker in our setting is an arbitrary program that interacts with a program of interest (the victim).
We require some interaction between the attacker and the victim that allows to start a transient execution phase.
Root causes of transient execution include branch and alias mispredictions, but also machine clears~\cite{ragab_rage_2021}.
For example, to exploit \spectre-v1 in~\autoref{fig:pht} it is sufficient that the attacker has access to the input \varidx to mistrain the branch predictor.
We also assume both the attacker and the victim share the same cache, which the attacker analyzes during the decoding phase to retrieve the secret.
The attacker can access the cache after execution of any instruction and does not need to wait for termination of the victim program.

The goal of the attacker is to break software isolation by reading a secret from address \secret{} outside its sandbox boundary.
While the attacker cannot directly access \secret{}, they can trick the victim into leaking sensitive information.
We consider data leakage both under normal and speculative execution.
Our attacker observes the address of executed \load instructions to see if any match \secret{}.
This is a common leakage model for speculative isolation~\cite{sok:spectre}.

To model the effects of speculative execution, our semantics can mispredict the outcome of all branch instructions and the address of all memory instructions in the victim.
This is the worst-case scenario in terms of leakage regardless of how attackers poison predictors.

\subsection{Beyond Software Isolation}
\label{sec:beyond-isolation}

While we focus on software isolation, our semantics can also serve as a building block for other properties. 
We show in Sections~\ref{sec:semantics} and~\ref{sec:cat-models} that the proposed semantics captures behaviors that are unobservable from the architectural point of view.
Guarnieri et al.~\cite{speculative-contracts} recently proposed a framework to specify hardware-software contracts and guarantee non-interference-style properties, such as constant time.
In this framework, contracts are formed of an execution mode (the semantics) and an observation mode (capturing the threat model). 
While the CAT models we present in the following can serve as an execution mode, defining an observation mode in axiomatic semantics is an open problem and left for future work.

\section{Speculative Control Flow}
\label{sec:semantics}

In this section, we show how we model control flow (\autoref{sec:cf}) and its speculation (\autoref{sec:scf}) axiomatically to detect \spectre-v1 (also known as \spectre-\textsc{PHT}).
We also discuss the speculation window (\autoref{sec:win}) and how to mitigate the attack (\autoref{sec:mit1}).

\subsection{Traditional Control Flow}
\label{sec:cf}

In a traditional model of computation, instructions are fetched, executed and retired in order (see~\autoref{sec:in-order}).
In this setting, the set $\executed$ (see~\autoref{sec:exec}) is entirely defined by the value of conditions in jump instructions.
The following cases relate instructions $\lb, {\lb'} $ along the same path.
  If $\lb$ immediately follows $\lb'$ in the program order and $\lb'$ is not a conditional jump targeting $\lb$, then $\lb$ can only execute if $\lb'$ does:
  $$\textit{If }\lb'+1 = \lb \land \lb' \neq \beqz r,\lb'' \textit{, then } \cf{\lb} \Rightarrow \cf{\lb'}$$
  When $\lb$ is the target of the direct jump $\lb'$, then $\lb$ can only execute if $\lb'$ does:
  $$\textit{If }\lb' = \jmp \lb \textit{, then } \cf{\lb} \Rightarrow \cf{\lb'}$$
  If $\lb'$ is both the immediate predecessor of $\lb$ and a conditional jump $\beqz r,\lb''$ targeting a different label, the dependency requires that $r \neq 0$, otherwise the jump would be taken:
  $$\textit{If }\lb'+1 = \lb \land \lb' = \beqz r,\lb'', \textit{then } \cf{\lb} \Rightarrow {({\cf{\lb'}} \land \rval{e_{\lb'}})}$$
  Here, we interpret $\rval{e_{\lb'}}$ as a boolean to denote $\rval{e_{\lb'}} \neq 0$.
  The dependency requires that $r = 0$ when $\lb'$ is a conditional jump $\beqz r,\lb$ targeting $\lb$. 
  This corresponds to the case where the jump is taken: 
  $$\textit{If }\lb' = \beqz r,\lb \textit{, then } \cf{\lb} \Rightarrow {({\cf{\lb'}} \land {\neg \rval{e_{\lb'}}})}$$

The predicate \textsc{pred} (see~\autoref{sec:language}) captures the static predecessors of an instruction.
In a given execution however, each event has a unique predecessor.
The set $\executed$ must satisfy this.
Combining the four cases above, a given instance of an instruction $\lb$ is executed if we have that
\begin{equation}
\label{eq:cf}
{\cf \lb} \Rightarrow {\bigvee\limits_{{\lb'} \in \pred \lb} \cfd{\lb}{\lb'}}
\end{equation}
where the \textsc{CFD} (\emph{control flow dependency}) predicate is defined as
{\small
\begin{equation}
\label{eq:cfd}
\hspace*{-1mm}\cfd{\lb}{\lb'} \define 
     \begin{cases}
       \cf{\lb'} & \text{if } \lb'+1 = \lb \land \lb' \neq \beqz r,\lb'' \\
       \cf{\lb'} & \text{if } \lb' = \jmp \lb \\
       {{\cf{\lb'}} \land \rval{e_{\lb'}}} & \text{if } \lb'+1 = \lb \land \lb' = \beqz r,\lb'' \\
       {{\cf{\lb'}} \land {\neg \rval{e_{\lb'}}}} & \text{if } \lb' = \beqz r,\lb \\
     \end{cases}
\end{equation}}
Note that (\ref{eq:cfd}) does not need to consider the case where $\lb'+1 = \lb \land \lb' = \jmp \lb''$ because such instructions are not part of $\pred \lb$.

As we mentioned before, the program in~\autoref{fig:pht} cannot access out-of-bounds memory because, before accessing \varA, \varidx is compared to the size of the array.
To confirm this claim, let us analyze the two possible control flow paths of this program.
If $\varidx < \varA.\texttt{size}$, the body of the if statement is executed.
Following~(\ref{eq:cf}), this corresponds to executing every single instruction, i.e. $\executed = \{ e_1, \dots, e_7\}$.
For this execution, (\ref{eq:cfd}) requires that $\rval{e_2} \not = 0$, otherwise the jump would have been taken and $e_3$--$e_6 \not \in \executed$.
As $\rval{e_2} \not = 0$, we have $r_1 < \varA.\texttt{size}$ and $\varA + r_1$ is in bounds.
This means the dashed \relname{rf} relation is not possible because $e_s: \secret$ and $e_4$ access different addresses ($(e_s,e_4) \not \in \relname{loc}$) and the reads-from relation requires $\relname{rf} \subseteq \relname{loc}$.
The second path has $\rval{e_2} = 0$.
In this execution the jump is taken and the control flow is redirected from $3$ to $7$, i.e. $\executed = \{ e_1, e_2, e_3, e_7\}$.
Since only $e_1$ is a read event and the addresses of $\varidx$ and $\secret$ are different, no \load can read from out-of-memory.
We conclude this execution does not read from \secret\ either and the whole program is safe.

\subsection{Speculative Control Flow}
\label{sec:scf}

Modern processors implement branch speculation. 
Suppose a conditional jump is fetched  but the value of its condition is not yet known.
Instead of stalling, the processor makes a prediction on which branch will be taken and continues executing speculatively.
Our semantics needs to consider the effects of the branch predictor and speculative execution.
Instructions following a correct prediction are eventually committed; mispredictions lead to transient executions that are eventually rolled back.

Let $\commited \subseteq \executed$ represent instructions that are eventually committed.
This set captures the case where instructions are executed \emph{(i)} under normal semantics or, \emph{(ii)} speculatively, but under a correct prediction.
Transient instructions are represented by the set $\transient \subseteq \executed$.
We consider as executed any instruction that is either committed or transiently executed and have that
$\executed = \commited \uplus \transient$.

For each event $e_\lb$ representing a conditional jump at $\lb$ we use proposition $\cp \lb$ to represent that the predicted branch direction was correct or the value of the condition known. 
Following the always mispredict semantics~\cite{GuarnieriKMRS20} (which has been shown sufficient to obtain security guarantees w.r.t. all branch predictors), we leave $\cp \lb$ unconstrained.

In~\autoref{sec:cf} we formalized the properties $\executed$ must fulfill for traditional semantics with predicates (\ref{eq:cf}) and (\ref{eq:cfd}).
We extend this now to capture the effects of the branch predictor.
The first two cases of (\ref{eq:cfd}) neither involve conditional jumps nor branch predictors and thus remain unchanged.
The two cases below are possible when instructions execute along the correct control flow (i.e., correct prediction or known value of the condition) and at least one of them is a conditional jump.
  If $\lb'$ is not only the immediate predecessor of $\lb$, but also a conditional jump targeting a different label, the dependency requires that $r \neq 0$ (otherwise the jump would been taken) and \emph{that the branch predictor is correct}:
  \begin{gather*}
  \textit{If }\lb'+1 = \lb \land \lb' = \beqz r,\lb'',\\
  \textit{then } \com{\lb} \Rightarrow ({\com{\lb'} \land \rval{e_{\lb'}}} \land \cp{\lb'})
  \end{gather*}
  If $\lb'$ is a conditional jump targeting $\lb$, the dependency requires that $r = 0$ and that \emph{the branch direction is correctly predicted}. This corresponds to the case where the jump is taken:
  \begin{gather*}
  \textit{If }\lb' = \beqz r,\lb,\\
  \textit{then } \com{\lb} \Rightarrow ({\com{\lb'} \land {\neg \rval{e_{\lb'}}}} \land \cp{\lb'})
  \end{gather*}

In the presence of branch predictors, committed instructions must follow the correct control flow:
\begin{equation}
{\com \lb} \Rightarrow {\bigvee\limits_{{\lb'} \in \pred \lb} \cfd{\lb}{\lb'}} \nonumber
\end{equation}
The control flow dependency definition from (\ref{eq:cfd}) needs to be updated with the two cases from above as follows:
{\small
\[   
\cfd{\lb}{\lb'} \define \!
     \begin{cases}
       \com{\lb'\!\!} & \hspace*{-2.5mm}\text{if } \lb' \! + \! 1 = \lb \land \lb' \!\neq \beqz r,\lb'' \\
       \com{\lb'\!\!} & \hspace*{-2.5mm}\text{if } \lb' = \jmp \lb \\
       {\com{\lb'\!\!} \!\land\! \rval{e_{\lb'}}} \land {\red \cp{\lb'}} & \hspace*{-2.5mm}\text{if } \lb' \! + \! 1 = \lb \land \lb' \!= \beqz r,\lb'' \\
       {\com{\lb'\!\!} \!\land\! {\neg \rval{e_{\lb'}}}} \land {\red \cp{\lb'}} & \hspace*{-2.5mm}\text{if } \lb' = \beqz r,\lb \\
     \end{cases}
\]}

We capture dependency between two instructions with transient execution in a similar way.
There are four possible cases; the first two cases are analogous to the traditional control flow case.
The differences with the remaining two cases are the following:
  If $\lb'$ is a conditional jump immediately preceding $\lb$ but targeting a different label, then $\lb$ executing transiently implies that $\lb'$ was executed (\emph{transiently or not}) and the jump should have been taken \emph{but the branch predictor made the wrong guess} and thus altered the control flow:
  \begin{gather*}
    \textit{If }\lb'+1 = \lb \land \lb' = \beqz r,\lb'',\\
    \textit{then } \scf{\lb} \Rightarrow ({{\cf{\lb'}} \land \neg \rval{e_{\lb'}}} \land {\neg \cp{\lb'}})
  \end{gather*}
  If $\lb$ is the target of the conditional jump $\lb'$, then $\lb$ executing transiently implies $\lb'$ was executed \emph{under any semantics} and the jump was taken due to a \emph{wrong prediction}:
  \begin{gather*}
  \textit{If }\lb' = \beqz r,\lb,\\
  \textit{then } \scf{\lb} \Rightarrow ({\cf{\lb'} \land {\rval{e_{\lb'}}}} \land \neg \cp{\lb'})
  \end{gather*}
We capture \emph{speculative control flow dependencies} with the constraint
\begin{equation}
\label{eq:scf}
{\scf \lb} \Rightarrow {\bigvee\limits_{{\lb'} \in \pred \lb} \scfd{\lb}{\lb'}}
\end{equation}
where the speculative control flow dependency \textsc{SCFD} is defined as
{\small
\[   
\scfd{l}{l'} \define \!
     \begin{cases}
       \scf{\lb'\!\!} & \hspace*{-2.5mm}\text{if } \lb' \! + \! 1 \!=\! \lb \land \lb' \! \neq \! \beqz r,\lb'' \\
       \scf{\lb'\!\!} & \hspace*{-2.5mm}\text{if } \lb' = \jmp \lb \\
       {{\cf{\lb'\!\!}} \! \land \! \neg \rval{e_{\lb'}}} \! \land \! {\neg \cp{\lb'}} & \hspace*{-2.5mm}\text{if } \lb' \! + \! 1\! = \! \lb \land \lb' \! = \! \beqz r,\lb'' \\
       {{\cf{\lb'\!\!}} \! \land \! {\rval{e_{\lb'}}}} \! \land \! {\neg \cp{\lb'}} & \hspace*{-2.5mm} \text{if } \lb' = \beqz r,\lb \\
     \end{cases}
\]}

Let us analyze the behavior of the program in~\autoref{fig:pht} in the presence of speculative execution.
The two execution paths from~\autoref{sec:cf} are still possible if $\cp{3}$ is true, i.e., if the branch predictor is correct.
Additionally, the program has two executions where the branch predictor is wrong and some instructions execute transiently:
\begin{eqnarray}
\label{sol3}
\neg \rval{e_2}, \neg \cp 3, \commited = \{ e_1, e_2, e_3 \}, \transient = \{ e_4, e_5, e_6, e_7 \} \\
\rval{e_2}, \neg \cp 3, \commited = \{ e_1, e_2, e_3 \}, \transient = \{ e_7 \} \nonumber
\end{eqnarray}
Note that (\ref{sol3}) has $\neg \rval{e_2}$ and thus $\varidx \geq \varA.\texttt{size}$.
Because $\varidx$ is not in bounds, it is possible that $\varA + r_1 = \secret$.
In this scenario, both $e_s: \secret$ and $e_4$ access the same address ($(e_s,e_4) \in \relname{loc}$) and thus the \relname{rf} edge is possible, showing that the secret can be read and the program is vulnerable to \spectre-v1.

\subsection{Speculation Window}
\label{sec:win}

Speculative execution attacks are only effective if there is a large enough speculation window for the transient execution phase.
Thus, we need not only to consider the effects of the branch predictor, but also the speculation window it creates.
Let \window be the size of the speculation window created by a branch prediction.
The theoretical upper limit of instructions that can be transiently executed during this window is given by the size of the reorder buffer.
Given a relation \relname{r}, the definition below computes the pairs that are related by composing \relname{r} with itself at most $k$ times:
$$\relname{r}^{\leq k} \define
\left\{
	\begin{array}{ll}
		\relname{r} & \mbox{if } k = 0 \\
		\relname{r};\relname{r}^{\leq k-1} & \mbox{otherwise}
	\end{array}
\right.
$$
We model that an instruction can only execute transiently during the speculation window using the following constraint to restrict the set $\transient$:
$\relname{po} \subseteq \relname{po} ; ([\transient] ; \relname{po})^{\leq \window-1}.$
The constraint imposes that the number of \relname{po}-consecutive transiently executed events is smaller than \window.
Note that $\transient$ can still have more than \window elements if several mispredictions occur and the corresponding transient events are not \relname{po}-consecutive.
The constraint also directly allows \emph{nested speculations} within the same window, which is challenging to achieve with operational semantics.

\subsection{Mitigating \spectre-\textsc{v1}}
\label{sec:mit1}

Serializing instructions can be used to stop the speculation and mitigate \spectre-v1.
We model this by enforcing that fences cannot be executed transiently:
\begin{equation}
\label{eq:fence}
\bigwedge\limits_{\lb \; : \; \fence} e_\lb \not \in \transient
\end{equation}
Together, constraints (\ref{eq:scf}) and (\ref{eq:fence}) imply that instructions following a \fence cannot execute transiently unless a new speculation is started by another conditional jump.
For example, adding a \fence after instruction $3$ in~\autoref{fig:pht} forbids the execution in (\ref{sol3}) even in the presence of speculation.
This shows that our semantics does not only allow to detect \spectre-v1 attacks, but also to prove that common mitigations work.

\section{Instruction Reordering}
\label{sec:cat-models}

In this section, we show how to model the logical reordering of instructions (e.g., due to aliasing or value forwarding), the reason underlying \spectre-v4. We begin with explaining in-order execution as a baseline~(\autoref{sec:in-order}) and then present our model for store-to-load forwarding (\autoref{sec:stl}) and possible defenses~(\autoref{sec:mitigating_sv4}). Next, we introduce our model for predictive store forwarding (\autoref{sec:sct}).
Finally, we show how CAT naturally handles the interaction between speculation, concurrency, and weak memory models (\autoref{sec:concurrency}) due to the composability of axiomatic models (\autoref{sec:composability}).

\subsection{In-order Execution}
\label{sec:in-order}

\begin{figure}[t]
\centering
\scalebox{.8}{

\begin{tikzpicture}

  \node (n1) {$e_1: \load r_0, \varA.\texttt{size}$};
  \node (n2) [below of=n1] {$e_2: \load r_1, \varidx$};
  \node (n3) [below of=n2] {$e_3: \store \varidx, r_1 \& (r_0-1)$};
  \node (n4) [below of=n3] {$e_4: \load r_2, \varidx$};

  \node (init)[left=1cm of n3] {$e_0: init$};

  \node (n5) [below of=n4] {$e_5: \load r_3, \varA + r_2$};
  \node (n6) [below of=n5] {$e_6: \load r_4, \varB + r_3$};
  \node (n7) [below of=n6] {$e_7: \load r_5, \vartemp$};
  \node (n8) [below of=n7] {$e_8: \store \vartemp, r_4 \& r_5$};

  \node(C) [left=.8cm of n1]{
  \begin{lstlisting}[style=CStyle]
idx = idx & (A.size - 1);
temp &= B[A[idx]];
  \end{lstlisting}
  };

  \path
	(n1) edge[->]			node[left] {\relname{po}} 		(n2)
	(n2) edge[->]			node[left] {\relname{po}} 		(n3)
	(n3) edge[->]			node[left] {\relname{po}} 		(n4)
	(n4) edge[->]			node[left] {\relname{po}} 		(n5)
	(n5) edge[->]			node[left] {\relname{po}} 		(n6)
	(n6) edge[->]			node[left] {\relname{po}} 		(n7)
	(n7) edge[->]			node[left] {\relname{po}} 		(n8);
  
  \path
	(n3.south east) edge[->, bend right=-40]			node[below] {\relname{rf}}		(n4.east);  

  \path
	(init) edge[->, bend left=-20]			node[below] {\relname{rf}}		(n5.north west);  

		\node [draw, rectangle, rounded corners, text=black, inner xsep=10pt, inner ysep=5pt] (a) at (-5,-6) {
		\scalebox{1}{{\small
			\begin{tabular}{p{.2\textwidth}}
			$\relname{com} = {{\relname{co} \cup \relname{rf}} \cup (\relname{rf}^{-1};\relname{co})}$ \\
			\hline
			\centering
			${\color{cobalt}\relname{acyclic}}\,\,\relname{com}  \cup \relname{po}$
			\end{tabular}}}
	}; 
	\node[fancytitle, right=10pt, font=\scriptsize,yshift=.5ex, fill=black, text=white] at (a.north west) {In-Order};

\end{tikzpicture}}
\caption{Safe array access due to index masking.}
\label{fig:stl-safe}
\end{figure}
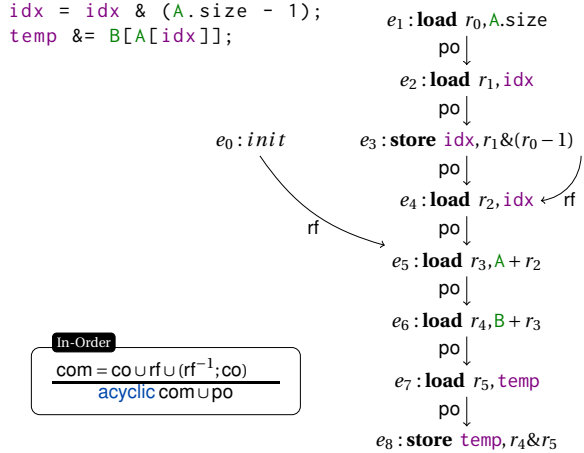

In the simplest model of execution, the processor fetches an instruction, stalls until its operands are available, and finally executes.
This model of computation, called \emph{in-order execution}, is straightforward to understand and analyze: instructions are executed one-by-one following the program order.
Consider the program in~\autoref{fig:stl-safe}. 
Since \varidx is masked using \varA.\texttt{size}, accessing array \varA is safe.
This safe behavior is captured by the candidate execution where event $e_5$ reads from the initial values of the array and there are no out-of-bounds accesses.
Since this program has only one control flow path (regardless of whether the processor supports speculative control flow or not), this is the only possible execution of the program under in-order semantics.

Our CAT model of the in-order semantics is shown in~\autoref{fig:stl-safe}.
As usual, \relname{rf} edges only relate write-read pairs accessing the same address.
The model defines a causal dependency $\relname{com} \cup \relname{po}$ and forces it to be acyclic (otherwise instructions could not be scheduled to satisfy their dependencies).
From $\relname{rf}^{-1};\relname{co}$ we have that if a read $r$ gets its value from a write $w$ ($\relname{rf}^{-1}(r,w)$), then any other write $w'$ coming after $w$ in the coherence order ($\relname{co}(w,w')$) must come after $r$ ($\relname{rf}^{-1};\relname{co}(r,w')$), otherwise $r$ would get its value from $w'$ instead.
Setting \relname{com} to be acyclic imposes that there is a single view of how instructions hit memory.
Since instructions are executed in order, \relname{po} must be part of the causal dependency.
All these imply event $e_4$ must read from the last \store to that address (here $e_3$), otherwise there would be a cycle.
Having $\relname{rf}(e_3,e_4)$ enforces that $r_2$ gets the masked value of \varidx.
Because of the masking, we have $r_2 < \varA.\texttt{size}$.
Finally, since the address of instruction $e_5$ is in-bounds, $e_5$ can only read the initial value of the array.

Instruction $e_3$ cannot directly read the initial value of \varidx (represented here by the $e_0: \mathit{init}$ event).
Reading the initial (not masked) value of $\varidx$, it would be possible for $e_5$ to access the secret.
The combination of \relname{rf} and \relname{co} given in~\autoref{fig:stl-unsafe} forms the candidate execution representing this unsafe scenario, but it forms a cycle of dependencies $e_4 \stackrel{\relname{\scriptsize rf}^{-1}}{\longrightarrow} e_0 \stackrel{\relname{\scriptsize co}}{\longrightarrow} e_3 \stackrel{\relname{\scriptsize po}}{\longrightarrow} e_4$.
Since the CAT model forbids cycles involving those relations, this unsafe scenario is not possible and we can conclude the program is safe under in-order semantics.

\subsection{Store-to-load Forwarding}
\label{sec:stl}

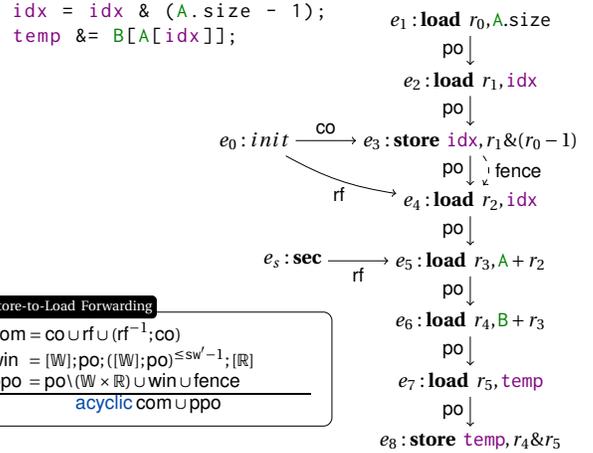
\begin{figure}[t]
\centering
\scalebox{.8}{

\begin{tikzpicture}

  \node (n1) {$e_1: \load r_0, \varA.\texttt{size}$};
  \node (n2) [below of=n1] {$e_2: \load r_1, \varidx$};
  \node (n3) [below of=n2] {$e_3: \store \varidx, r_1 \& (r_0-1)$};
  \node (n4) [below of=n3] {$e_4: \load r_2, \varidx$};

  \node (init)[left=1cm of n3] {$e_0: init$};

  \node (n5) [below of=n4] {$e_5: \load r_3, \varA + r_2$};
  \node (n6) [below of=n5] {$e_6: \load r_4, \varB + r_3$};
  \node (n7) [below of=n6] {$e_7: \load r_5, \vartemp$};
  \node (n8) [below of=n7] {$e_8: \store \vartemp, r_4 \& r_5$};

  \node (secret)[left=1cm of n5] {$e_s: \secret$};
  
  \node(C) [left=.8cm of n1]{
  \begin{lstlisting}[style=CStyle]
idx = idx & (A.size - 1);
temp &= B[A[idx]];
  \end{lstlisting}
  };

  \path
	(n1) edge[->]			node[left] {\relname{po}} 		(n2)
	(n2) edge[->]			node[left] {\relname{po}} 		(n3)
	(n3) edge[->]			node[left] {\relname{po}} 		(n4)
	(n4) edge[->]			node[left] {\relname{po}} 		(n5)
	(n5) edge[->]			node[left] {\relname{po}} 		(n6)
	(n6) edge[->]			node[left] {\relname{po}} 		(n7)
	(n7) edge[->]			node[left] {\relname{po}} 		(n8);
  
  \path
	(secret) edge[->]			node[below] {\relname{rf}}		(n5);  
  \path
	(init) edge[->]			node[above] {\relname{co}}		(n3);  
  \path
	(init) edge[->, bend left=-10]			node[below] {\relname{rf}}		(n4);  

  \path
	(n3) edge[->, bend right=-40, dashed]			node[right] {\relname{fence}}		(n4);  


		\node [draw, rectangle, rounded corners, text=black, inner xsep=10pt, inner ysep=5pt] (a) at (-5.35,-5.8) {
		\scalebox{1}{{\small
			\begin{tabular}{p{.29\textwidth}}
			$\relname{com} = {{\relname{co} \cup \relname{rf}} \cup (\relname{rf}^{-1};\relname{co})}$ \\
			$\relname{win} \hspace{1.1mm} = {[\stores] ; \relname{po} ; ([\stores] ; \relname{po})^{\leq \window' -1} ; [\loads]}$ \\
			$\relname{ppo} \hspace{.6mm} = {{\relname{po} \backslash (\stores \times \loads)} \cup \relname{win} \cup \relname{fence}}$ \\
			\hline
			\centering
			${\color{cobalt}\relname{acyclic}}\,\, \relname{com}  \cup \relname{ppo}$
			\end{tabular}}}
	}; 
	   \node[fancytitle, right=10pt, font=\scriptsize,yshift=.5ex, fill=black, text=white] at (a.north west) {Store-to-Load Forwarding};

\end{tikzpicture}}
\caption{\spectre-v4 -- impossible under \emph{in-order} semantics, but possible under \emph{store-to-load forwarding} semantics.}
\label{fig:stl-unsafe}
\end{figure}

Modern processors 
use a combination of speculative and out-of-order execution optimizations,
so some of the guarantees from in-order execution do not hold.
Although not observable at the architectural level, the microarchitecture allows more executions, which might leave traces in the microarchitectural state.
In~\autoref{fig:stl-unsafe}, if the address used by $e_3$ is not yet known, the processor might predict that $e_3$ and $e_4$ will not alias and speculatively execute $e_4$ before $e_3$.
This is the basis of \spectre-\textsc{STL} (one instance of \spectre-v4).
Our CAT model for \emph{store-to-load forwarding} still imposes a single view of how instructions hit memory (\relname{com} is acyclic), but uses a weaker notion of preserved program order (\relname{ppo}) than in-order semantics (which uses the whole \relname{po}) and thus allows the \spectre-\textsc{STL} behavior.
There are three possible scenarios where the order of events is preserved by \relname{ppo}:
\begin{itemize}
  \item Events are in program order, but they are not a write-read pair.
  \item They are \enquote{far away} in the program order in such a way that events in between fill the store buffer.
  \item The corresponding instructions are separated by a \fence and thus the events are related by a \relname{fence} edge.
\end{itemize}
The relation ${\relname{po} \backslash (\stores \times \loads)}$ tells us that some read events can be speculatively executed before a \relname{po}-previous write.
This is allowed, e.g., if the processor knows or predicts the addresses do not alias.
Apart from write-read pairs, our CAT model preserves every other pair of events in program order since they are irrelevant to model store-to-load forwarding.

However, not all write-read pairs can be reordered.
Once a write event has been committed (it is no longer in the store buffer), its address is known, alias speculation is not possible, and the store cannot be passed over.
The size of the store buffer defines a speculation window that can be modeled in CAT by tracking the number of write events between the potentially reordered pair.
To model the speculation window, we preserve the program order between a write-read pair if the number of write events between the pair fills the buffer, forcing to retire and commit the write event.
We model this in CAT using the relation $\relname{win} = {[\stores] ; \relname{po} ; ([\stores] ; \relname{po})^{\leq \window' -1} ; [\loads]}$ where $\window'$ is the size of the store buffer, with $\window'  \geq 1$.
Note that $\window'$ can differ from the speculation window $\window$ created by the branch predictor.
For $\relname{r}^{\leq k}$ to be well defined, $k \geq 0$ must hold.
Relations $[\stores]$ and $[\loads]$ are the identity relation restricted to write and read events respectively, i.e., $[\stores] = \{ (e,e) \mid e \in \stores \}$ and $[\loads] = \{ (e,e) \mid e \in \loads \}$.
Using $[\stores]$ on the left of the composition guarantees that the first event in every pair is a write (those not to be reordered).
Analogously, using $[\loads]$ on the right forces the second event to be a read.
The inner part $\relname{po} ; ([\stores] ; \relname{po})^{\leq \window' -1}$ represents the writes between the potentially reordered pair that fit in the buffer.
For example, if the size of the store buffer is two, \relname{win} reduces to ${[\stores] ; \relname{po} ; [\stores] ; \relname{po} ; [\stores] ; \relname{po} ; [\loads]}$, which relates all write-read pairs in \relname{po} for which there are at most two write events in between.

Since \fence instructions stop speculation, the \relname{ppo} also preserves the order of events coming from instructions for which there is a fence in between. 
If we add a \fence between instructions $3$ and $4$ in the \masm program for~\autoref{fig:stl-unsafe}, this results in the dashed fence edge being part of the graph.

The preserved program order \relname{ppo} is defined as the union of the three relations described above.
Let us see how this semantics affect the possible behavior of the program.
The pair $(e_3,e_4)$ is not part of \relname{ppo}: it is a write-read pair from \relname{po} which can be reordered because the buffer might not be full  
and there are no fences in between (even if the buffer was full, the processor can decide to commit some of the buffered events, making room for $e_3$).
Since it is possible to reorder $e_3$ and $e_4$ and because the initial value of \varidx is controlled by the attacker, the value loaded to $r_2$ can be bigger than \varA.\texttt{size}.
The address of $e_5$ depends on $r_2$, allowing to potentially read out of bounds and access the secret.
The candidate execution shown in~\autoref{fig:stl-unsafe} represents this scenario and it is consistent according to our CAT model, i.e., our semantics models \spectre-\textsc{STL}.

\begin{figure}[t]
\centering
\scalebox{.8}{

\begin{tikzpicture}

  \node (n1) {$e_1: \load r_0, \varA.\texttt{size}$};
  \node (n2) [below of=n1] {$e_2: \load r_1, \varidx$};
  \node (n3) [below of=n2] {$e_3: r_2 \leftarrow r_1 \& (r_0-1)$};

  \node (init)[left=1cm of n3] {$e_0: init$};

  \node (n5) [below of=n3] {$e_4: \load r_3, \varA + r_2$};
  \node (n6) [below of=n5] {$e_5: \load r_4, \varB + r_3$};
  \node (n7) [below of=n6] {$e_6: \load r_5, \vartemp$};
  \node (n8) [below of=n7] {$e_7: \store \vartemp, r_4 \& r_5$};

  \node(C) [left=.8cm of n1]{
\begin{lstlisting}[style=CStyle]
register ridx asm ("r2");
ridx = idx & (A.size - 1);
temp &= B[A[idx]];
\end{lstlisting}
  };

  \path
	(n1) edge[->]		node[left] {\relname{po}} 		(n2)
	(n2) edge[->]		node[left] {\relname{po}} 		(n3)
	(n3) edge[->]		node[left] {\relname{po}} 		(n5)
	(n5) edge[->]		node[left] {\relname{po}} 		(n6)
	(n6) edge[->]		node[left] {\relname{po}} 		(n7)
	(n7) edge[->]		node[left] {\relname{po}} 		(n8);
  
  \path
	(init) edge[->, bend left=-10]	node[below left] {\relname{rf}}		(n5.west);  

\end{tikzpicture}}
\caption{\spectre-v4 -- store replaced by register assignment (safe).}
\label{fig:stl-reg}
\end{figure}
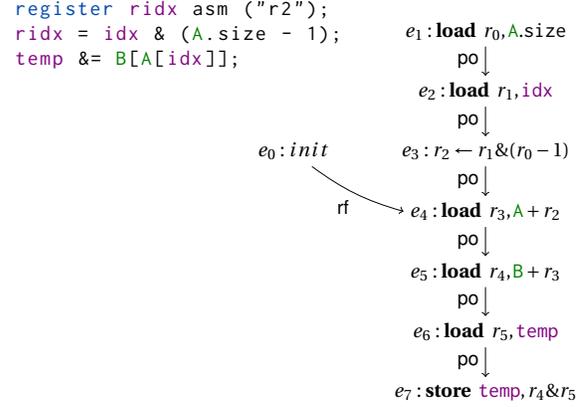

\subsection{Mitigating \spectre-\textsc{STL}}
\label{sec:mitigating_sv4}

At the software level, there are at least two possible ways to mitigate \spectre-\textsc{STL}.
One alternative is to add a \fence instruction resulting in the dashed fence edge in~\autoref{fig:stl-unsafe}.
Since pairs of events related by \relname{fence} are part of the \relname{ppo} order, adding this edge results in the cycle $e_4 \stackrel{\relname{\scriptsize rf}^{-1}}{\longrightarrow} e_0 \stackrel{\relname{\scriptsize co}}{\longrightarrow} e_3 \stackrel{\relname{\scriptsize fence}}{\longrightarrow} e_4$, which is not allowed by the CAT model.
This shows that the mitigation makes the program safe under our semantics.
Another alternative is given in~\autoref{fig:stl-reg}.
Here, the compiler is instructed to store the masked value in a register, resulting in a different assembly program. 
Since the attack relies on bypassing a store to memory, but the masking is now done by a local computation using only registers, our semantics guarantees that $e_4$ will get the masked value in $r_2$ and thus the access to \varA will be in bounds.

\subsection{Predictive Store Forwarding}
\label{sec:sct}

The \spectre variant from~\autoref{sec:stl} exploits the fact that processors may predict that the addresses of two instructions do not alias.
Some processors implement the opposite, allowing them to predict that two addresses alias even if they finally do not match.
This type of prediction was used as the basis of a theoretical new version of \spectre-v4~\cite{CauligiDGTSRB20,GuancialeBD20} which we denote \spectre-\textsc{PSF}.
AMD recently confirmed that its Zen 3 process is vulnerable to this attack~\cite{amd-sec2}. 
We present a CAT model allowing this kind of prediction and modeling such an attack.

\begin{figure}[t]
\centering
\scalebox{.8}{

\begin{tikzpicture}

  \node (n1)  {$e_1: r_1 \leftarrow \varidx < \varC.\texttt{size}$};
  \node (n2) [below of=n1] {$e_2: \beqz r_1, \bot$};
  \node (n3) [below of=n2]{$e_3: \store \varC+0, 64$};
  \node (n4) [below of=n3]{$e_4: \load r_2, \varC + \varidx$};
  \node (n5) [below of=n4]{$e_5: \load r_3, r_2 * \varidx$};

  \node (n6) [below of=n5] {$e_6: \load r_4, \varA + r_3$};
  \node (n7) [below of=n6] {$e_7: \load r_5, \varB + r_4$};
  \node (n8) [below of=n7] {$e_8: \load r_6, \vartemp$};
  \node (n9) [below of=n8] {$e_{9}: \store \vartemp, r_5 \& r_6$};


  \node (secret)[left=1cm of n6] {$e_s: \secret$};
  
  \node(C) [left=.5cm of n2]{
  \begin{lstlisting}[style=CStyle]
uint8_t A[16];
uint8_t C[2] = { 0,0 };
if (idx < C.size) {
  C[0] = 64;
  temp &= B[A[C[idx] * idx]];
}
  \end{lstlisting}
  };

  \path
	(n1) edge[->]			node[left] {\relname{po}} 		(n2)
	(n2) edge[->]			node[left] {\relname{po}} 		(n3)
	(n3) edge[->]			node[left] {\relname{po}} 		(n4)
	(n4) edge[->]			node[left] {\relname{po}} 		(n5)
	(n5) edge[->]			node[left] {\relname{po}} 		(n6)
	(n6) edge[->]			node[left] {\relname{po}} 		(n7)
	(n7) edge[->]			node[left] {\relname{po}} 		(n8)
	(n8) edge[->]			node[left] {\relname{po}} 		(n9);
  
  \path
	(n3) edge[->, bend right=-40, dashed]			node[right] {\relname{fence}}		(n4);  
  \path
	(n3.east) edge[->, bend left=50]	node[right] {\relname{srf}}		(n4.east);  
  \path
	(secret) edge[->]			node[below] {\relname{rf}}		(n6);  

		\node [draw, rectangle, rounded corners, text=black, inner xsep=10pt, inner ysep=5pt] (a) at (-5.35,-6.5) {
		\scalebox{1}{{\small
			\begin{tabular}{p{.23\textwidth}}
			$\relname{scom} = {{\relname{co} \cup \relname{srf}} \cup (\relname{srf}^{-1};\relname{co})}$ \\
			\hline
			\centering
			${\color{cobalt}\relname{acyclic}}\,\, \relname{scom}  \cup \relname{po}$
			\end{tabular}}}
	}; 
	   \node[fancytitle, right=10pt, font=\scriptsize,yshift=.5ex, fill=black, text=white] at (a.north west) {Predictive Store Forwarding};

\end{tikzpicture}}
\caption{\spectre-v4 -- predictive store forwarding (unsafe).}
\label{fig:sct}
\end{figure}
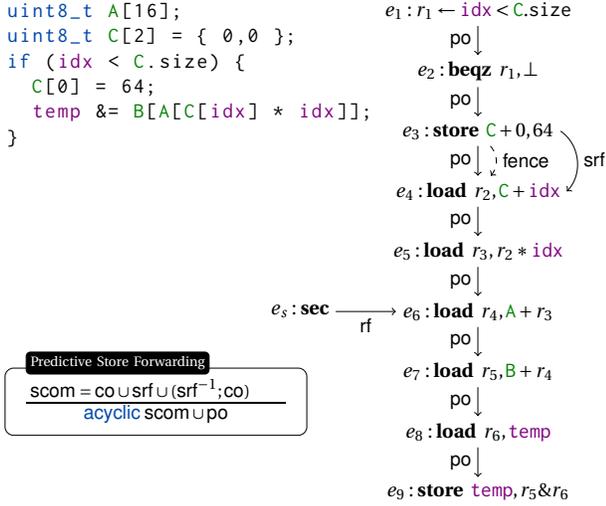

The unsafe scenario is shown in~\autoref{fig:sct}.
Note that this program accesses array \varC using an attacker-controlled index \varidx.
Even if the index is compared with the size of \varC, the program is vulnerable to \spectre-v1 as we have seen in~\autoref{sec:scf}.
Unfortunately, even if we use the mitigation from~\autoref{sec:mit1} to guarantee that $e_4$ is not transiently executed, the program is still vulnerable in the presence of alias speculation.
Consider the scenario where \varidx is 1.
Despite having different offsets, the processor can speculate that $e_3$ and $e_4$ alias.
In this scenario, register $r_2$ gets the value 64 which is then multiplied by 1 (the value of \varidx) to compute the address of $e_5$.
Since the size of \varA is smaller than 64, instruction $e_6$ performs an out-of-bounds access allowing to read the secret.

Our CAT model for \emph{predictive store forwarding} is given in~\autoref{fig:sct}.
The formalization closely follows the in-order semantics with the exception that instead of using relation \relname{rf}, we use a new relation that supports alias speculation.
While events related by \relname{rf} must access the same address, \relname{srf} relaxes this and only imposes that the addresses of those two instructions must alias: 
\begin{gather*}
\relname{srf} \subseteq (\stores \times \loads) \cap \relname{alias} \\ 
\forall r \in \loads : {\exists! w \in \stores : {\relname{srf}(w,r)}} \\
\relname{srf}(w,r) \Rightarrow \rval{w} = \rval{r}
\end{gather*}
\noindent
We leave the $\relname{alias}$ relation unconstrained to support arbitrary predictors, i.e., $\relname{alias} \define \{ (m_1,m_2) \mid m_1,m_2 \in \memory \}$.

The candidate execution from~\autoref{fig:sct} fulfills the assertion of our predictive store forwarding CAT model showing that the program is vulnerable to \spectre-\textsc{PSF}.

Since \fence instructions stop speculation, we must enforce that if two events are related by \relname{srf} and there is a \fence between the corresponding instructions, then the events must access the same address, i.e., ${\relname{srf} \cap \relname{fence}} \subseteq {\relname{loc}}$.
Note that if the dashed \relname{fence} edge in~\autoref{fig:sct} is part of the graph, the edge $e_3 \stackrel{\relname{\scriptsize srf}}{\rightarrow} e_4$ would not be allowed because $\varC + 0 \not = \varC + \varidx$ (in this scenario \varidx = 1), forcing $e_4$ to read its value from $e_0: \mathit{init}$ and forbidding $e_6$ to access \secret.

\subsection{Concurrency and Weak Memory Models}
\label{sec:concurrency}

While all the attacks we described previously have an inherent notion of concurrency (the attacker resides in a sandbox different from the victim), so far we only considered single-threaded victims.
Besides branch misprediction, the Intel Optimization Reference Manual~\cite{intel-opt} lists \emph{machine clear} (i.e., data misprediction) as a category of \emph{bad speculation}.
One possible cause of such machine clears is memory ordering in the presence of concurrency.
Intel and AMD processors implement the \emph{Total-Store-Order} memory model in which all cores see operations in program order except in one case: a \store followed by a \load on a different address may be reordered. 
This is because cores use local buffers to hide the latency of \store operations.

\begin{figure}[t]
\centering
\scalebox{.8}{

\begin{tikzpicture}

  \node (n1) {$e_1: \load r_0, x$};
  \node (n2) [below of=n1] {$e_2: \load r_1, y$};
  \node (n3) [below of=n2] {$e_3: \load r_2, \varA + r_0*(r_0-r_1)$};

  \node (init)[left=1cm of n2] {$e_0: init$};

  \node (n6) [right=1.6cm of n1] {$e_6: \store r_2,y$};
  \node (n7) [below of=n6] {$e_7: \store r_3,x$};
  \node (n5) [above of=n6] {$e_5: r_3 \leftarrow 1$};
  \node (n4) [above of=n5] {$e_4: r_2 \leftarrow 1$};

  \node (secret)[left=.45cm of n3] {$e_s: \secret$};

  \node(C) at (-4,0.5){
  \begin{lstlisting}[style=CStyle]
uint8_t A[1];

thread_1:
r0 = x;
r1 = y;
temp &= B[A[r1-r0]];

thread_2:
y = 1;
x = 1;
  \end{lstlisting}
  };

  \path
	(n1) edge[->]			node[left] {\relname{po}} 		(n2)
	(n2) edge[->]			node[left] {\relname{po}} 		(n3)
	(n4) edge[->]			node[left] {\relname{po}} 		(n5)
	(n5) edge[->]			node[left] {\relname{po}} 		(n6)
	(n6) edge[->]			node[left] {\relname{po}} 		(n7);
  
  \path
	(init) edge[->]			node[above] {\relname{rf}}		(n2.west);  

  \path
	(n7.west) edge[->]			node[below left, rotate=35] {\relname{rf}$^{-1}$;\relname{co}}		(n1.east);  

  \path
	(n2.east) edge[->]			node[above left, rotate=-35] {\relname{rfe}}		(n6.west);  

  \path
	(secret) edge[->]			node[above] {\relname{rf}}		(n3);  

		\node [draw, rectangle, rounded corners, text=black, inner xsep=10pt, inner ysep=5pt] (a) at (-3,-4) {
		\scalebox{1}{{\small
			\begin{tabular}{p{.29\textwidth}}
			$\relname{com} = {{\relname{co} \cup \relname{rf}} \cup (\relname{rf}^{-1};\relname{co})}$ \\
			$\relname{com-tso} = {{\relname{co} \cup \relname{rfe}} \cup (\relname{rf}^{-1};\relname{co})}$ \\
			$\relname{po-tso} = ({{{\relname{po}} \cap {({(\loads \times \memory)} \cup {(\stores \times \stores)}})}) \cup {\relname{fence}}}$ \\
			\hline
			\centering
			${\color{cobalt}\relname{acyclic}}\,\,\,\relname{com}  \cup {(\relname{po} \cap \relname{add})}$ \\
			${\color{cobalt}\relname{acyclic}}\,\,\relname{com-tso}  \cup \relname{po-tso}$
			\end{tabular}}}
	}; 
	\node[fancytitle, right=10pt, font=\scriptsize,yshift=.5ex, fill=black, text=white] at (a.north west) {\tso};

\end{tikzpicture}}
\caption{Transient execution due to invalid memory ordering.}
\label{fig:concurrency}
\end{figure}
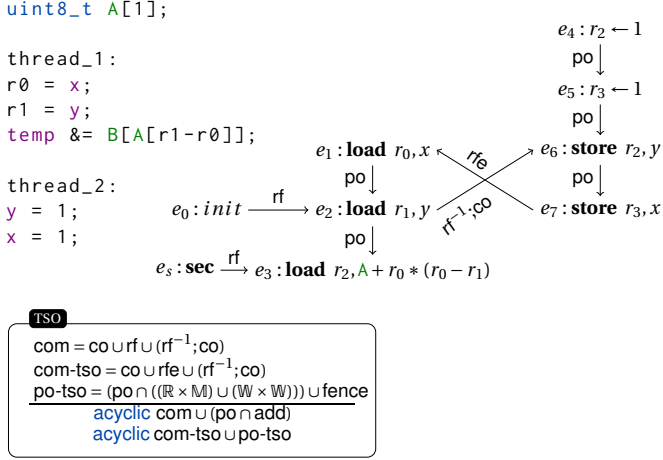

The CAT model for \tso~\cite{SarkarSNORBMA09} is given in~\autoref{fig:concurrency}.
The assertion ${\color{cobalt}\relname{acyclic}}\,\,\relname{com}  \cup {(\relname{po} \cap \relname{loc})}$ enforces local consistency within the same core.
Due to the use of buffers, only the \emph{external reads-from relation} is part of the global view.
Here, \relname{rfe} is the subset of \relname{rf} such that the events from the pair belong to different cores.
Relation \relname{po-tso} keeps all pairs in \relname{po} except write-read pairs (which from a global point of view might be reordered), unless there is a \fence instruction between them.

Consider the concurrent program in~\autoref{fig:concurrency} running on the \tso memory model.
Here we assume \masm is extended with threads running on different cores: a program is composed by a set of threads and a thread is a sequence of instructions.
The only impact this change has in the CAT semantics is that we must guarantee \relname{po} only relates events within the same thread.
The program shows a \emph{message passing} pattern where if \texttt{thread\_1} observes \varx = 1, then it should also observe \vary = 1.
Since the \tso semantics guarantees $r_0*(r_0-r_1)$ is always 0, the access to {\color{mygreen} \texttt{A}} is safe.
The CAT model captures this since it forbids the (unsafe) candidate execution where $r_0 = 1 \land r_1 = 0$ due to the cycle $e_1 \stackrel{\relname{\scriptsize po}}{\longrightarrow} e_2 \stackrel{\relname{\scriptsize rf$^{-1};\relname{\scriptsize co}$}}{\longrightarrow} e_6 \stackrel{\relname{\scriptsize po}}{\longrightarrow} e_7 \stackrel{\relname{\scriptsize rfe}}{\longrightarrow} e_1$.

Ragab et al.~\cite{ragab_rage_2021} noticed that if $e_1$ is a slow \load (due to a cache miss) and $e_2$ a fast one (cache hit), event $e_2$ can be transiently executed before $e_1$.
Suppose that while the \load from \varx is pending, both stores from \texttt{thread\_2} are executed as represented by the candidate execution in~\autoref{fig:concurrency}.
If this is the case, the coherence controller notifies that the values of \varx and \vary have changed, leading to a \emph{memory ordering machine clear} which guarantees that the behavior is not observable at the architectural level.
Abusing this kind of behavior is non-trivial due to the strict synchronization requirements, however there exists a speculation window which could be use to mount an attack.

We can use CAT to capture this speculation window by replacing $(\loads \times \memory)$ with $(\loads \times \stores)$ and adding \relname{addr} to the union in the definition of \relname{po-tso}.
These changes allow some read-read pairs to be reordered.
The first assertion still guarantees the reordering is not possible if both reads access the same address; and by adding \relname{addr} to \relname{po-tso}, we forbid to revert the order when the address of the second read depends on the value loaded by the first one.
Since $e_1, e_2$ neither access the same address nor have an address dependency, with this modification, $(e_1,e_2) \not \in \relname{po-tso}$ and the candidate execution becomes consistent.
It is then possible that $r_0*(r_0-r_1) \not = 0$ and thus, $e_3$ can read the secret if $\varA + 1 = \secret$.

\subsection{Composability of Axiomatic Models}
\label{sec:composability}

One of the main advantages of axiomatic models is that they are easily composable.
Any of the control flow semantics in~\autoref{sec:semantics} can be combined with any of the CAT models in this section.
For instance, if an attack requires to mistrain both the branch predictor and the memory alias predictor, this would be detected by combining the \textsc{scfd} definition in~\autoref{sec:scf} and the Store-to-Load Forwarding CAT model in~\autoref{fig:stl-unsafe}.

\newcommand\usedtimeout{90 }

\section{Implementation}
\label{sec:framework}

We use Bounded Model Checking (BMC) parametrized by our axiomatic semantics defined as CAT models for verifying software isolation. 
All predicates from Sections~\ref{sec:semantics} and~\ref{sec:cat-models} can be encoded using first order logic over the domain of booleans and integers.
In fact, a compact BMC encoding to test reachability using the CAT language with traditional control flow was developed in~\cite{porthos,dartagnan-cav}.
We implemented \zombmc\footnote{\url{https://github.com/unibw-patch/Kaibyo}}, a prototype tool to test software isolation, as an extension of \dartagnan\footnote{\url{https://github.com/hernanponcedeleon/Dat3M}}.
Apart from the definition of the new CAT models, we added an x86 parser and the speculative control flow encoding from~\autoref{sec:scf}, we modified the property being checked (from reachability to software isolation), and we implemented the new \relname{srf} relation.

\zombmc takes as inputs a program written in \xes assembly, a CAT model, an unrolling bound $k$ and an address \secret.
It generates a formula which is satisfiable if and only if there is a consistent execution (according to the CAT model and where loops where unrolled up to $k$ iterations) where some read event reads from $e_s: \secret$.
As any other BMC technique, we only analyze bounded (finite) executions of programs containing loops.
Thus, \zombmc only provides security guarantees up to the given bound.
Nevertheless, any transient execution gadget found by the tool can be exploited under the conditions laid out in~\autoref{sec:threatmodel}.

\section{Evaluation}
\label{sec:evaluation}

We use \zombmc to evaluate our approach and answer the following research questions:
\begin{description}[labelwidth=\widthof{\bfseries RQ1:}, leftmargin=!]
\item [RQ1:] \emph{Do our semantics cover known attacks?}
\item [RQ2:] \emph{Do our semantics prove effectiveness of proposed countermeasures?}
\item [RQ3:] \emph{What effort is required to support new semantics in the analysis?}
\item [RQ4:] \emph{How complex are the generated formulae for state-of-the-art SMT solvers?}
\end{description}
In the following, we first discuss our experimental setup (\autoref{sec:setup}) and then evaluate the ability of our models to capture \spectre variants (\autoref{sec:prec}), the flexibility of the tool to support different semantics (\autoref{sec:flex}), and the performance of SMT solvers on the generated formulae (\autoref{sec:perf}).

\subsection{Experimental Setup}
\label{sec:setup}

We compare the results of \zombmc against \spectector~\cite{GuarnieriKMRS20} and \binsec~\cite{DanielBR20} using the following benchmarks
\begin{description}[labelwidth=\widthof{\bfseries (PHT)}, leftmargin=!]
  \item[(PHT)] Fifteen benchmarks by Kocher~\cite{variants} exploiting branch prediction.
  \item[(STL)] Thirteen benchmarks from the \binsec repository~\cite{spectrev4-litmus} exploiting store-to-load forwarding.
  \item[(PSF)] The example program from~\autoref{fig:sct} (adapted from~\cite{amd-sec2}) exploiting predictive store forwarding.
\end{description}
All benchmarks are written in C and compiled using GCC 8.3.0. 
The results of our evaluation are given in~\autoref{table:all}.
The expected result w.r.t software isolation when no mitigation is used (\none column) is either \safe ({\mygreen +}) or \unsafe ({\red -}).
For each benchmark there is a variant (\fen column) using a \fence instruction to stop branch or alias speculation; all such variants are \safe.
A \gtick\ entry means the tool returns the corrected expected result. 
We show \danger\ if the tool cannot analyze the program and detail the reasons below.

\subsection{Precision of the CAT Models}
\label{sec:prec}

\begin{figure}[h]
  \input{Figures/table-all}
  \caption{Evaluation with different \spectre behaviors. Expected result when no mitigation is used is \safe ({\mygreen +}) or \unsafe ({\red --}). The result of the tool is correct \gtick\ or it cannot analyze the program \danger.}
  \label{table:all}
\end{figure}

All three tools support branch prediction and correctly report all PHT benchmarks as vulnerable to \spectre-v1.
Both \zombmc and \spectector also prove that adding fences after each conditional jump stops speculation. 
\binsec has no support for \fence instructions and thus it cannot analyze the benchmarks using the mitigation. 
\zombmc returns \safe for all benchmarks except PHT-05 which contains an input dependent loop.
In the presence of input dependent loops, BMC techniques can find violations (see \none column), but they cannot prove programs correct, because loops cannot be fully unrolled.
While the same limitation applies to \spectector and \binsec (which are based on symbolic execution), they return \safe even if the analysis is incomplete.
\zombmc returns \unknown when it cannot find a violation up to the given bound; at the same time, it cannot prove the bound is large enough to explore all executions, hence the \danger\ entry in the table.

\spectector has no support for store-to-load forwarding and thus it cannot analyze any of the STL benchmarks.
A stack overflow occurs when running \zombmc to generate the formula for STL-09 which requires a large (\textgreater 200) unrolling bound. 
\zombmc proves that adding fences between reordered pairs makes the programs \safe.
\binsec cannot analyze the mitigated benchmarks due to not supporting fences.
Both \zombmc and \binsec prove that instructing the compiler to use registers instead of \store instructions (as in~\autoref{fig:stl-reg}) makes STL-03 and STL-12 \safe.
Both tools agree in all remaining results except for STL-13.
The reason is that, as we describe below, the tools do not consider the same threat model.

\begin{figure*}[]
\centering
  \scalebox{.8}{

\begin{tikzpicture}

  \node(C) []{
  \begin{lstlisting}[style=CStyle]
uint8_t load_value(uint32_t idx) {
  register uint32_t ridx asm ("edx");
  ridx = idx & (A.size - 1);
  uint8_t to_leak = A[ridx];
  return to_leak;
}

void case_13(uint32_t idx) {
  register uint8_t to_leak asm ("edx");
  to_leak = load_value(idx);
  temp &= B[to_leak * 512];  
}
  \end{lstlisting}
};

\node[rectangle,minimum height=0.9cm,minimum width=3cm,draw,color=myred, thick] at (16.45,-1.04) {};
\node[rectangle,minimum height=0.45cm,minimum width=2.5cm,draw,color=myred, thick] at (9.65,-.4) {};


  \node(ass) at (14.8,.3) {
  \begin{lstlisting}[style=x86Style]
load_value:
	sub	esp, 16
	mov	eax, A.size
	sub	eax, 1
	and	eax, [esp+20]
	mov	edx, eax
	mov	eax, edx
	movzx	eax, A[eax]
	mov	[esp+15], al
	movzx	eax, [esp+15]
	add	esp, 16
	ret
  \end{lstlisting}
};

  \node(ass) at (8,.3) {
  \begin{lstlisting}[style=x86Style]
  case_13:
	push	[esp+4]
	call	load_value
	add	esp, 4
	mov	edx, eax
	mov	eax, edx
	movzx	eax, al
	movzx	edx, B[eax]
	movzx	eax, temp
	and	eax, edx
	mov	temp, al
	ret
  \end{lstlisting}
};

\end{tikzpicture}}
  \caption{A variant of \spectre-v4 from~\cite{spectrev4-litmus} written in C (left) and compiled to \xes with GCC 8.3.0 (right).}
  \label{fig:case13}
\end{figure*}

The original C code of STL-13 and the corresponding \xes assembly are given in~\autoref{fig:case13}.
After masking the index and accessing \texttt{{\varA}[\texttt{\color{violet}ridx}]}, function \texttt{\color{violet}load\_value} moves the loaded value to the stack and back before returning.
If the instruction \texttt{mov eax, [esp+15]} happens before the previous \store (i.e. they are reordered), then it can read from uninitialized memory.
If the value from the initialized memory is bigger than the size of \varB, then the access in \texttt{\color{violet}case\_13} to \texttt{{\varB}[\texttt{eax}]} is out of bounds and can access the secret.
This access brings the secret to the cache and this, according to our threat model from~\autoref{sec:threatmodel}, might have security consequences; thus \zombmc reports it as \unsafe.
However, even if the secret is loaded into \texttt{edx}, it does not influence the control flow (e.g., conditional branches) or memory addresses (e.g., offsets into arrays).
Therefore the program does not violate speculative constant time, the property \binsec verifies.

\zombmc is able to detect that PSF-01 from~\autoref{fig:sct} is vulnerable to \spectre-v1.
It also reports that even if adding a \fence after the conditional jumps stops the control flow speculation, the program remains vulnerable due to predictive store forwarding.
Finally, by inserting a unique \fence between instructions $3$ and $4$, to stop both branch and alias speculation, it proves the program \safe.
To the best of our knowledge, \zombmc is the only tool having support for predictive store forwarding.

The results above answer {\bfseries RQ1} and {\bfseries RQ2} by showing that CAT can be used to define semantics covering \spectre-v1 and v4 and proving that common mitigations work.

\subsection{Flexibility of the Analysis}
\label{sec:flex}

To answer {\bfseries RQ3}, we report on our development efforts to support all vulnerabilities discussed in this paper.
\zombmc is based on the tool \dartagnan~\cite{dartagnan-svcomp2021} which implements the traditional control flow encoding from~\autoref{sec:cf} and supports the core of the CAT language from~\autoref{fig:model}.
\dartagnan verifies Boogie code~\cite{leino2008this} and most of our development effort was spent in parsing \xes assembly and encoding the stack.
Extending the tool to support speculative control flow required around 100 lines of Java code.
Adding support to store-to-load forwarding and memory order machine clear only required to develop the CAT models from~\autoref{fig:stl-unsafe} and the variation of \tso described in~\autoref{sec:concurrency}.
The tool then automatically detected both \unsafe behaviors.
Detecting the attack based on predicative store forwarding required implementing the axioms of the new relation \relname{srf}.
The SMT encoding of \relname{srf} enforces that both events should alias instead of access the same address, as in \relname{rf}.
To implement the \relname{alias} relation we used an unconstrained boolean variable for every pair of memory events.

\subsection{SMT Performance}
\label{sec:perf}

To answer {\bfseries RQ4}, we compare the performance of four different SMT solvers (\zthree, \cvcfour, \yices and \mathsat) on the generated formulae from~\autoref{table:all}.
The results are given in~\autoref{fig:smt}.
Times are shown in seconds using a logarithmic scale.
We used a \usedtimeout min timeout.
For entries \danger\ in the table, we treat the result as a timeout.
Since SMT solvers tend to perform differently for \sat and \unsat instances, we divided the results in two.
The top of the figure shows the results for the original benchmarks (no mitigation). 
All formulae are \sat except STL-03 and STL-12 (no formula is generated for STL-09 due to a stack overflow).
The bottom of the figure present the results for the benchmarks in the \fen column which are all \unsat instances.

For the \sat instances \yices generally wins, with \zthree and \mathsat being competitive.
\cvcfour shows always the worst performance for these \sat formulae, sometimes three orders of magnitude slower than the other solvers.
However, \cvcfour shows the overall best performance for the \unsat instances.
In particular, for benchmarks STL and PSF where reasoning about instruction reordering is needed, most solvers timeout.
\cvcfour is able to solve all such benchmarks except STL-01 for which \mathsat has the best performance.

The bottom of the figure shows a clear difference in the search space (which the solver needs to completely explore since these instances are \unsat) of PHT w.r.t STL and PSF.
While there are four possible outcomes for branch prediction (following the true or false branch combined with correctly or incorrectly predicted), alias misprediction allows a \load to draw data from \emph{any} prior {\bf store}, making the search space much bigger.
This suggests that it is beneficial for tools to have a portfolio of SMT backends; this is possible using libraries like JavaSMT, for instance~\cite{BaierBF20,KarpenkovF016}.


\pgfplotstableread[col sep=comma,]{Figures/z3-times.csv}\ztimes
\pgfplotstableread[col sep=comma,]{Figures/yices-smt2-times.csv}\ytimes
\pgfplotstableread[col sep=comma,]{Figures/cvc4-times.csv}\ctimes
\pgfplotstableread[col sep=comma,]{Figures/mathsat-times.csv}\mtimes
\pgfplotstablegetcolsof{\ytimes}
\pgfmathtruncatemacro{\NoOfCols}{\pgfplotsretval-1}

\begin{figure*}
\begin{tikzpicture}
    \begin{semilogyaxis}[
        width=2.15*\axisdefaultwidth,
        height=.62*\axisdefaultheight,
        ybar=0pt,
        ymin=0.01,
        ylabel={Time (secs)},
        ytick={1e-2,1e-1,1e0,1e1,1e2,1e3,1e4},
        y tick label style={font=\scriptsize},
        yminorticks=false,
        xtick=data,
        x tick label style={font=\scriptsize},
        xticklabels={
        \scriptsize PHT-01 ({\red --}), \scriptsize PHT-02 ({\red --}), \scriptsize PHT-03 ({\red --}), \scriptsize PHT-04 ({\red --}), \scriptsize PHT-05 ({\red --}), \scriptsize PHT-06 ({\red --}), \scriptsize PHT-07 ({\red --}), \scriptsize PHT-08 ({\red --}), \scriptsize PHT-09 ({\red --}), \scriptsize PHT-10 ({\red --}), \scriptsize PHT-11 ({\red --}), \scriptsize PHT-12 ({\red --}), \scriptsize PHT-13 ({\red --}), \scriptsize PHT-14 ({\red --}), \scriptsize PHT-15 ({\red --}), 
        \scriptsize STL-01 ({\red --}), \scriptsize STL-02 ({\red --}), \scriptsize STL-03 ({\mygreen +}), \scriptsize STL-04 ({\red --}), \scriptsize STL-05 ({\red --}), \scriptsize STL-06 ({\red --}), \scriptsize STL-07 ({\red --}), \scriptsize STL-08 ({\red --}), \danger \scriptsize STL-09 ({\mygreen +}), \scriptsize STL-10 ({\red --}), \scriptsize STL-11 ({\red --}), \scriptsize STL-12 ({\mygreen +}), \scriptsize STL-13 ({\red --}),
	\scriptsize PSF-01 ({\red --})
        },
        x tick label style={rotate=45,anchor=east},
        log origin=infty,
        bar width=0.2,
        enlarge x limits={abs=0.6},
	legend style={at={(0.35,0.85)},anchor=west}
    ]
	\addplot table [x expr=\coordindex,y index=1,col sep=comma,] {\ztimes};
	\addplot table [x expr=\coordindex,y index=1,col sep=comma,] {\ytimes};
	\addplot table [x expr=\coordindex,y index=1,col sep=comma,] {\ctimes};
	\addplot table [x expr=\coordindex,y index=1,col sep=comma,] {\mtimes};
	\pgfplotstablegetcolumnnamebyindex{1}\of{\ytimes}\to{\colname};
	\addplot[black!40,dashed,sharp plot,update limits=false] 
		coordinates {(-1,5400) (29,5400)}
		node[above] at (axis cs:10,500) {\color{black} \small Timeout: 90 min};    
    \end{semilogyaxis}
\end{tikzpicture}
\begin{tikzpicture}
    \begin{semilogyaxis}[
        width=2.15*\axisdefaultwidth,
        height=.62*\axisdefaultheight,
        ybar=0pt,
        ymin=0.01,
        ylabel={Time (secs)},
        ytick={1e-2,1e-1,1e0,1e1,1e2,1e3,1e4},
        y tick label style={font=\scriptsize},
        yminorticks=false,        
        xtick=data,
        x tick label style={font=\scriptsize},
        xticklabels={
        \scriptsize PHT-01 ({\mygreen +}), \scriptsize PHT-02 ({\mygreen +}), \scriptsize PHT-03 ({\mygreen +}), \scriptsize PHT-04 ({\mygreen +}), \danger \scriptsize PHT-05 ({\mygreen +}), \scriptsize PHT-06 ({\mygreen +}), \scriptsize PHT-07 ({\mygreen +}), \scriptsize PHT-08 ({\mygreen +}), \scriptsize PHT-09 ({\mygreen +}), \scriptsize PHT-10 ({\mygreen +}), \scriptsize PHT-11 ({\mygreen +}), \scriptsize PHT-12 ({\mygreen +}), \scriptsize PHT-13 ({\mygreen +}), \scriptsize PHT-14 ({\mygreen +}), \scriptsize PHT-15 ({\mygreen +}),
        \scriptsize STL-01 ({\mygreen +}), \scriptsize STL-02 ({\mygreen +}), \scriptsize STL-03 ({\mygreen +}), \scriptsize STL-04 ({\mygreen +}), \scriptsize STL-05 ({\mygreen +}), \scriptsize STL-06 ({\mygreen +}), \scriptsize STL-07 ({\mygreen +}), \scriptsize STL-08 ({\mygreen +}), \danger \scriptsize STL-09 ({\mygreen +}), \scriptsize STL-10 ({\mygreen +}), \scriptsize STL-11 ({\mygreen +}), \scriptsize STL-12 ({\mygreen +}), \scriptsize STL-13 ({\mygreen +}),
        \scriptsize PSF-01 ({\mygreen +})
        },
        x tick label style={rotate=45,anchor=east},
        log origin=infty,
        bar width=0.2,
        enlarge x limits={abs=0.6},
	legend style={at={(0.88,0.25)},anchor=west,nodes={scale=0.5, transform shape}}
    ]
	\addplot table [x expr=\coordindex,y index=2,col sep=comma,] {\ztimes};
	\addplot table [x expr=\coordindex,y index=2,col sep=comma,] {\ytimes};
	\addplot table [x expr=\coordindex,y index=2,col sep=comma,] {\ctimes};
	\addplot table [x expr=\coordindex,y index=2,col sep=comma,] {\mtimes};
	\pgfplotstablegetcolumnnamebyindex{2}\of{\ytimes}\to{\colname}
	\addplot[black!40,dashed,sharp plot,update limits=false] 
		coordinates {(-1,5400) (29,5400)}
		node[above] at (axis cs:10,500) {\color{black} \small Timeout: 90 min};    
	\addlegendentryexpanded{\zthree};
	\addlegendentryexpanded{\yices};
	\addlegendentryexpanded{\cvcfour};
	\addlegendentryexpanded{\mathsat};
    \end{semilogyaxis}
\end{tikzpicture}
\caption{Benchmarks from columns \none (top) and \fen (bottom). Labels show if the formula is \sat ({\red --}) or \unsat ({\mygreen +}). \zombmc could not fully unroll the program (PHT-05) or generate the formula (STL-09) for entries marked with \danger.}
\label{fig:smt}
\end{figure*}
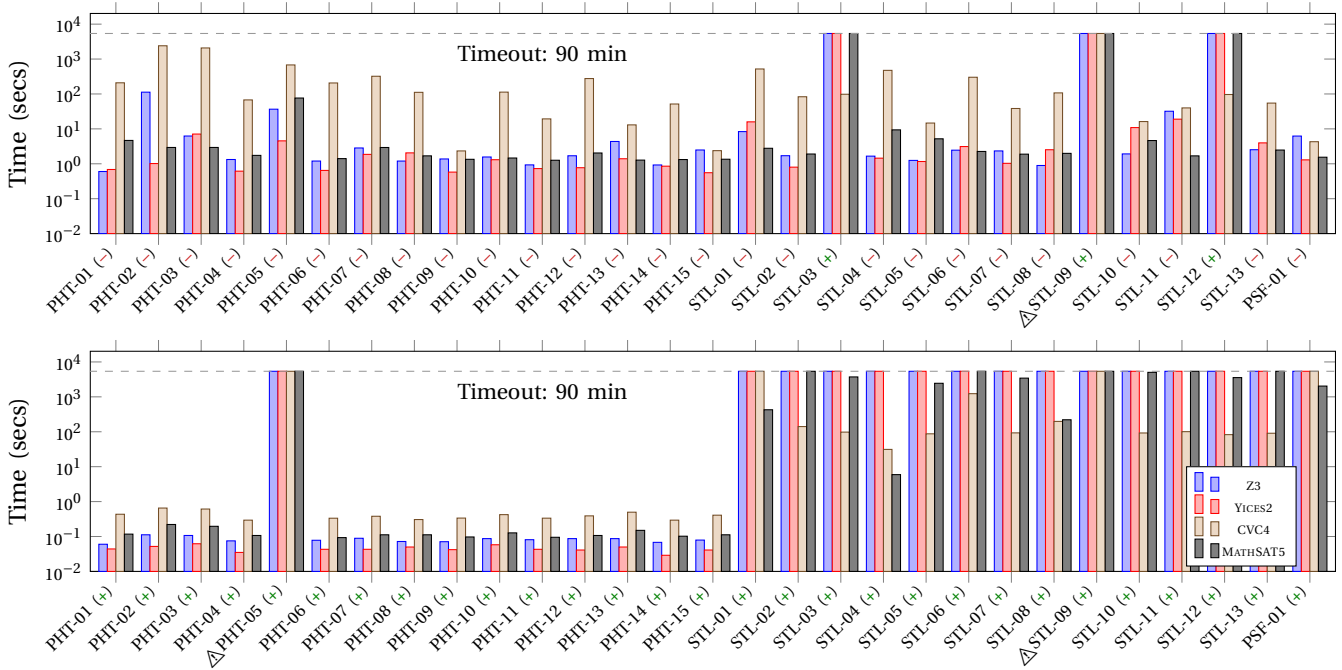

The solving times of \spectector and \binsec for each of the benchmarks in~\autoref{table:all} are between 0.1 and 10 seconds, meaning they are one or two orders of magnitude faster than \zombmc.
This is not surprising since those tools are specialized for concrete models and thus less flexible. 
Also, despite branch misprediction and instruction reordering, the benchmarks have few execution, favoring symbolic execution approaches (which use a \emph{simple SMT query for each execution}) over BMC ones (which use a \emph{complex SMT query for all executions}).

\section{Discussion}
\label{sec:discussion}

\zombmc analyzes programs w.r.t software isolation.
Currently, it cannot analyze non-interference-style properties, and
our evaluation shows the consequences of this.
\zombmc considers a program unsafe if it accesses the secret, even if this does not violate, e.g., constant time, because it neither influences the control flow nor the memory addresses. 
As we discussed in~\autoref{sec:beyond-isolation}, given a notion of observation for axiomatic models, our semantics could be used to also verify non-interference-style properties.
Since this requires reasoning about pairs of executions, doing so would affect the performance of our implementation, however.

\zombmc demonstrates that it is possible to translate axiomatic semantics into a concrete analysis tool, but it is a proof-of-concept implementation with clear limits to scalability.
Improving performance of tools based on axiomatic semantics is an active area of research; the weak memory model community recently made progress on scalable non-BMC tools based on axiomatic semantics~\cite{Kokologiannakis21}.

\section{Related Work}
\label{sec:related}

This section describes the work that has been done on security foundations since the disclosure of \spectre.
Formal microarchitectural models are capable of representing out-of-order and speculative behavior.
Side channels are modeled by different notions of observations which over-approximate the attacker capabilities and abstract from the memory subsystem and cache.
Security guarantees are formalized as properties comparing such observations under two different semantics: a reference execution model (generally in-order execution) and a target execution model (e.g., speculative and/or out-of-order executions).

Currently, the target execution model of Guarnieri et al.~\cite{GuarnieriKMRS20} only covers branch speculation and captures \spectre-v1 behaviors by the notion of \emph{speculative non-interference}.
\spectector is a symbolic execution tool for testing speculative non-interference.
To extend its semantics to other \spectre versions, one would have to adapt the notion of microarchitectural state.
Cauligi et al. extend constant time in the presence of speculation, leading to the new notion of \emph{speculative constant time}~\cite{CauligiDGTSRB20}.
Even though their semantics models most of the behaviors underlying \spectre (including indirect jumps, return stack buffers and predictive store forwarding), their analysis tool \pitchforktool does not support any of these.
Doing so \emph{\enquote{would require to generate a prohibitively large number of possible schedules}}. 
Guanciale et al.~\cite{GuancialeBD20} define an out-of-order execution model using microarchitectural instructions rather than the ISA.
They show traditional constant time (which is defined at the ISA level) is not secure enough \emph{even in the absence of speculation} and propose a constant time property for the out-or-order execution model.
The proposed semantics is very expressive and even allows to mispredict arbitrary values.
Unfortunately, there is no tool based on such semantics.

The approaches above can be formalized as \emph{hardware-software contracts} specifying which program executions an attacker can differentiate~\cite{speculative-contracts}.
On the one side, contracts provide security foundations to build tools that help software developers write microarchitecturally secure programs.
On the other side, for hardware developers, contracts are specifications that describe allowed microarchitectural effects without enforcing a specific implementation.
Contracts are 2-hypersafety properties~\cite{ClarksonS10} and thus require appropriate techniques to efficiently model pairs of traces.
RelSE is a promising approach to extend symbolic execution for analyzing security properties of two execution traces~\cite{binsec}.
In fact, an extension to RelSE has recently been proposed to test speculative constant time~\cite{DanielBR20}.

We were not the first to provide insights about the relation between \spectre attacks and the weak memory models which characterize modern hardware.
Disselkoen et al. use pomsets (where edges also represent dependencies) to model executions~\cite{DisselkoenJJR19}.
The novel aspect of this model is that events have preconditions which can captured failed branch predictions.
Their semantics cover \spectre-v1 and models out-of-order executions, but it is not clear if they capture \spectre-v4.
There is also no tool based on this semantics.
\checkmate~\cite{checkmate} uses graphs to capture the subtle orderings of hardware execution events when programs run on a microarchitecture.
Their ``micro-architecturally happens-before'' notion is based on \relname{po}, \relname{rf} and \relname{co}, but their executions are less abstract than ours: a single instruction is represented by many events modeling fetching, execution, commit and completion.
They even explicitly model when a value is brought to and flushed from the L1 cache.
One of the novelties of their work is the use of graphs to represent exploit patterns like \flushreload and \primeprove.
It remains an open question if CAT can be use to capture such patterns.
Closest to our work are Colvin and Winter~\cite{ColvinW19}.
They integrate speculative control flow within a more general verification framework using \impro, a language for reasoning about weak memory models.
While \impro can handle weak memory models such as \tso, \power and \arm~\cite{ColvinS18}, their paper focuses on sequential consistency instead of exploring the interaction between speculation and weak memory models as we did in~\autoref{sec:concurrency}.
Finally, CAT has seen far wider adoption based on the large body of literature~\cite{AlglaveMMPS18, AlglaveMT14, BattyDW16, PulteFDFSS18, SarkarSNORBMA09} and its use in industry~\cite{AlglaveDGHM21}.

\section{Conclusion}
\label{sec:conclusion}

We studied the use of axiomatically defined semantics in the presence of speculative and out-of-order execution, a domain that, contrary to its operational counterpart, had not been sufficiently explored.
We showed that CAT, a domain specific language initially developed to clarify the concurrency semantics of weak memory models, can also be used to analyze the consequences of microarchitectural optimizations.
Challenging aspects of speculative execution, such as speculation nesting, are naturally modeled using our framework.
Although our axiomatic BMC-based prototype is slower than its operational symbolic execution-based counterparts, the pluggable abstract semantics allow for a cleaner and more general implementation that can easily accommodate new models, as demonstrated by the ability to quickly add support for new attacks such as that based on memory ordering machine clears~\cite{ragab_rage_2021}.

\section*{Acknowledgments}

We would like to thank our anonymous shepherd and the anonymous reviewers for their comments and feedback. We would also like to thank Sébastien Bardin, Lesly-Ann Daniel, and Marco Guarnieri for valuable discussions on this work.

\bibliographystyle{plain}
\bibliography{../../common/bibliography}

\end{document}